\documentclass[fleqn,usenatbib]{mnras}

\usepackage{newtxtext,newtxmath}

\usepackage[T1]{fontenc}

\DeclareRobustCommand{\VAN}[3]{#2}
\let\VANthebibliography\thebibliography
\def\thebibliography{\DeclareRobustCommand{\VAN}[3]{##3}\VANthebibliography}

\usepackage{graphicx}	
\usepackage{amsmath}	
\usepackage{longtable}
\usepackage{booktabs}
\newcommand{\orcid}[1]{\href{https://orcid.org/#1}{\includegraphics[width=8pt]{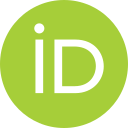}}}

\title[VAST-P1 Late-Time Supernovae Re-Brightening]{Late-Time Supernovae Radio Re-brightening in the VAST Pilot Survey}

\author[K. Rose et al.]{Kovi Rose \orcid{0000-0002-7329-3209}$^{1,2,3}$\thanks{E-mail: kovi.rose@sydney.edu.au},
Assaf Horesh \orcid{0000-0002-5936-1156}$^{1}$,
Tara Murphy \orcid{0000-0002-7329-3209}$^{2,4}$,
David ~L.\ Kaplan \orcid{0000-0002-7329-3209}$^{5}$,
Itai Sfaradi \orcid{0000-0002-7329-3209}$^{1}$,
\newauthor
Stuart D. Ryder \orcid{0000-0002-7329-3209}$^{6,7}$,
Robert J. Aloisi \orcid{0000-0002-7329-3209}$^{5,8}$,
Dougal Dobie \orcid{0000-0002-7329-3209}$^{2,4}$,
Laura Driessen \orcid{0000-0002-7329-3209}$^{2}$,
Rob Fender \orcid{0000-0002-7329-3209}$^{9}$,
\newauthor
David A. Green \orcid{0000-0002-7329-3209}$^{10}$,
James K. Leung \orcid{0000-0002-9415-3766}$^{11,12,1}$,
Emil Lenc \orcid{0000-0002-7329-3209}$^{3}$,
Hao Qiu \orcid{0000-0002-7329-3209}$^{13}$,
and David Williams-Baldwin \orcid{0000-0002-7329-3209}$^{14}$
\\
$^{1}$Racah Institute of Physics, The Hebrew University of Jerusalem, Jerusalem 91904, Israel\\
$^{2}$Sydney Institute for Astronomy, School of Physics, The University of Sydney, New South Wales 2006, Australia\\
$^{3}$CSIRO Astronomy and Space Science, PO Box 76, Epping, NSW 1710, Australia\\
$^{4}$ARC Centre of Excellence for Gravitational Wave Discovery (OzGrav), Hawthorn, VIC 3122, Australia\\
$^{5}$Center for Gravitation, Cosmology, and Astrophysics,
  Department of Physics, University of Wisconsin-Milwaukee, PO Box
  413, Milwaukee, WI, 53201, USA\\  
$^{6}$School of Mathematical and Physical Sciences, Macquarie University, NSW 2109, Australia\\
$^{7}$Astrophysics and Space Technologies Research Centre, Macquarie University, Sydney, NSW 2109, Australia\\
$^{8}$Department of Astronomy, University of Wisconsin-Madison, 475 North Charter Street, Madison, WI 53706, USA\\
$^{9}$Astrophysics, Department of Physics, University of Oxford, Keble Road, Oxford OX1 3RH, UK\\
$^{10}$Astrophysics Group, Cavendish Laboratory, 19 J. J. Thomson Ave., Cambridge CB3 0HE, UK\\
$^{11}$David A. Dunlap Department of Astronomy and Astrophysics, University of Toronto, 50 St. George Street, Toronto, ON M5S 3H4, Canada\\
$^{12}$Dunlap Institute for Astronomy and Astrophysics, University of Toronto, 50 St. George Street, Toronto, ON M5S 3H4, Canada\\
$^{13}$SKA Observatory, Jodrell Bank, Lower Withington, Macclesfield, Cheshire, SK11 9FT, UK\\
$^{14}$Jodrell Bank Centre for Astrophysics, School of Physics and Astronomy, The University of Manchester, Manchester, M13 9PL, UK\\
}

\date{Accepted XXX. Received YYY; in original form ZZZ}

\pubyear{2024}

\begin{document}
\label{firstpage}
\pagerange{\pageref{firstpage}--\pageref{lastpage}}
\maketitle

\begin{abstract}
We present our analysis of supernovae serendipitously found to be radio-bright several years after their optical discovery. We used recent observations from the Australian SKA Pathfinder taken as part of the pilot Variables and Slow Transients and Rapid ASKAP Continuum Survey programs. We identified $29$ objects by cross-matching sources from these ASKAP observations with known core-collapse supernovae below a declination of $+40^{\circ}$ and with a redshift of $z\leq0.15$. Our results focus on eight cases that show potential late-time radio emission. These supernovae exhibit significantly greater amounts of radio emission than expected from the standard model of a single shockwave propagating through a spherical circumstellar medium, with a constant density structure produced by regular stellar mass-loss. We also discuss how we can learn from future ASKAP surveys about the circumstellar environments and emission mechanisms of supernovae that undergo late-time radio re-brightening. This pilot work tested and confirmed the potential of the Variables and Slow Transients survey to discover and study late-time supernova emission.
\end{abstract}




\section{Introduction}
\label{sec:intro}

Massive stars shed mass via different mechanisms throughout their evolution. It has been suggested that some massive stars undergo increasingly violent eruptions during the final stages of stellar evolution, releasing large amounts of mass ranging from $10^{-7}$--$10^{-2}$ M$_\odot$ yr$^{-1}$
\citep[e.g.][and references therein]{2012ARA&A..50..107L,2014ARA&A..52..487S}. Stars with an initial mass of more than 8~M$_\odot$ are generally expected to explode in an energetic astronomical event known as a core-collapse supernova (CCSN). CCSNe occur when massive stars exhaust their nuclear fuel and collapse inwards due to overwhelming gravitational forces \citep{2003ApJ...591..288H}.

The radio emission that we detect from CCSNe arises from a fast-moving outflow of the SN ejecta -- which may be relativistic or sub-relativistic -- interacting with the circumstellar medium (CSM) and driving a shockwave that amplifies the magnetic field and accelerates electrons in the surrounding environment \citep{1982ApJ...259..302C}. In the post-shocked region, particles are accelerated to relativistic energies. 

The spectrum of these relativistic electrons, as described by \cite{1998ApJ...499..810C}, has a power-law distribution given by $N(E)=N_0E^{-\gamma}$, where $N$ is the number density of electrons, $N_0$ is a normalisation constant, and $\gamma$ is the spectral index of the relativistic electron energy density. A small fraction of the post-shock energy goes into enhancing magnetic fields,  which causes the gyration of relativistic electrons and produces the synchrotron emission \citep{1996ASPC...93..125C}. 

Some of these synchrotron photons can in turn be reabsorbed by nearby relativistic electrons in a process known as synchrotron self-absorption (SSA). Free--free absorption (FFA) -- generated from free electrons decelerating and scattering off charged ions -- can also modify the emitted radiation \citep[e.g.][]{2002ARA&A..40..387W}. FFA generally only dominates the emission at early times or in cases of slow-moving shocks \citep[e.g.][]{1998ApJ...499..810C,2024arXiv240307048M}.

Modelling the radio emission enables us to probe the circumstellar or interstellar medium (ISM), derive physical parameters of the shock, and reveal the progenitor star's mass-loss history. Examples of this sort of parameter space exploration can be seen in the analyses conducted by \cite{1998ApJ...499..810C}, \cite{2012ApJ...752...17W}, \cite{2020ApJ...903..132H}, \cite{2021ApJ...908...75B}, and others.

The standard model of a CSM with constant density structure is understood to be the result of constant mass-loss over thousands of years before the explosion. For relatively old supernovae (SNe) radio observations provide a unique set of tools to examine the progenitor mass-loss variations over long timescales. In contrast to the prompt emission usually detected within days or weeks of the optical discovery, the physical processes that generate comparable levels of radio emission many months or years after the supernova explodes are not as well understood. In most cases the emission gradually fades over time as the shock loses energy and the CSM becomes more diffuse.  However, there are instances where the radio emission is not monotonic, but brightens years after the SN.

Instances of such late-time radio re-brightening are rare but there are prominent cases like SN\,2007bg \citep{2013MNRAS.428.1207S}, SN\,2014C \citep{2017MNRAS.466.3648A}, and PTF11qcj \citep{2021ApJ...910...16P}. These SNe exhibit a double peak in their radio lightcurves, showing two clear periods of re-brightening roughly $10^{2}$--$10^{3}$ days after their optical discovery. Analyses of these SNe \citep{2013MNRAS.428.1207S, 2017MNRAS.466.3648A,2021ApJ...910...16P} support the interpretation of radio re-brightening caused by a propagating shock front interacting with two (or more) distinct CSM regions ejected prior to the explosion.

The rarity of late-time radio emission is partly due to the fact that most SNe are not followed up at late times after explosion. The same is true for Tidal Disruption Events (TDEs) which have also been found to exhibit late-time radio re-brightening \citep[e.g.][]{2022ApJ...933..176S,2024arXiv240712097A}.

It is anticipated that wide-field all-sky radio surveys will produce more detections of re-brightening events.
For example, the Karl G. Jansky Very Large Array \citep[VLA;][]{perley2011} Sky Survey \citep[VLASS;][]{Lacy_2020} -- which covers $\sim82$ per cent of the sky ($33,885$\,deg$^2$) at $2$--$4$\,GHz -- has identified new transients and late-time SNe re-brightening events \citep[e.g.][]{2021ApJ...923L..24S}. VLASS has a typical root mean square (RMS) noise sensitivity of $0.12$\,mJy beam$^{-1}$. The Australian SKA Pathfinder \citep[ASKAP;][]{2021PASA...38....9H} telescope is an ideal instrument for detecting radio transients in the southern hemisphere, with a large field-of-view and a typical RMS noise sensitivity $\sim 0.25$\,mJy beam$^{-1}$ for a $15$\,minute observation.

In this paper we present our search for CCSNe that exhibit late-time radio re-brightening. In Section \ref{sec:  Observations} we summarise the relevant ASKAP surveys and technical specifications, as well as the data reduction from follow-up observations. Section \ref{sec: Methodology} describes the candidate selection process and our lightcurve modelling and analysis. In Section \ref{sec: Results} we present our findings for all of the sources of interest. Finally, in Section \ref{sec: Discussion}, we discuss the implications of our results regarding ASKAP and the detection of similar late-time emission from CCSNe.

\section{Observations}
\label{sec: Observations}

\subsection{ASKAP Survey Observations}
\label{subsec: ASKAP Observations}

ASKAP is located at Inyarrimanha Ilgari Bundara, the CSIRO’s Murchison radio astronomy observatory in Western Australia. It is a radio telescope array comprised of $36$ parabolic antennae. Each dish is $12$\,m in diameter and has a phased array feed (PAF) that enables a large $30$\,deg$^2$ field-of-view. We used data from the Rapid ASKAP Continuum Survey \citep[RACS\footnote{\url{https://research.csiro.au/racs}};][]{2020PASA...37...48M} and the Variable and Slow Transients \citep[VAST\footnote{\url{https://vast-survey.org}};][]{2013PASA...30....6M} survey. The observations from both surveys were reduced with the \texttt{ASKAPSoft} data reduction pipeline \citep{2012SPIE.8500E..0LC}, which uses daily observations of PKS B1934$-$638 for band-pass, flux-density scale, and on-axis leakage calibration. After initial calibration the pipeline runs imaging and self-calibration on the data from each of the beams; see \cite{2021PASA...38....9H} for further details. The  \texttt{Selavy} source-finding package \citep{2012MNRAS.421.3242W} is implemented in \texttt{ASKAPSoft}. It produces catalogues of both extended and point sources by fitting Gaussian components to neighbouring pixels with a signal above a set detection threshold.

The first RACS data release \citep[RACS-low;][]{2021PASA...38...58H} covers the sky south of $+41^{\circ}$ declination ($\sim34\,240$\,deg$^2$), observed between April 2019 and June 2020 with $\sim$15~minute observations at a central frequency of $887.5$\,MHz and an angular resolution of $15$ arcsec. We used the RACS-low source catalogue as the main survey for our cross-matching. The VAST Phase I Pilot Survey \citep[VAST-P1;][]{2021PASA...38...54M} has a total sky footprint of $5\,131$\,deg$^2$ with $12$ minute observations conducted between August 2019 and August 2020 at the same observing frequency. We used the VAST-P1 catalogue and other publicly available ASKAP observations as additional lightcurve epochs, including data from the ongoing full VAST survey which covers a $\sim9\,500$\,deg$^2$ sky region at roughly a bimonthly cadence. To obtain the ASKAP flux uncertainties used in this work we took the quadrature sum of the \texttt{Selavy} peak flux error, image RMS, and a $6$ per cent flux scaling error. 

\subsection{AMI-LA\& ATCA Follow-Up Observations}
\label{subsec: Data Reduction}

We conducted follow-up radio observations of selected sources with the Arcminute Microkelvin Imager - Large Array \citep[AMI-LA;][]{zwart_2008,hickish_2018} and the Australia Telescope Compact Array \citep[ATCA;][]{2011MNRAS.416..832W}. The results of these observations are presented in Section \ref{sec: Results} and the measurements are available in the supplementary material.

The data from AMI-LA were initially flagged and calibrated using $\tt{reduce \_ dc}$, a customised data reduction software package \citep{perrott_2013}. We used the \texttt{CASA} (Common Astronomy Software Applications) package \citep{2007ASPC..376..127M} for additional flagging and imaging of the AMI-LA data, as well as the reduction of archival observations retrieved from the VLA Data Archive\footnote{\url{https://archive.nrao.edu/archive/advquery.jsp}}. Observations from ATCA were reduced using the \texttt{Miriad} software package \citep{1995ASPC...77..433S}.

AMI-LA observes with a $5$\,GHz bandwidth around a central frequency of $15.5$\,GHz. We conducted two $4$\,hour observations of SN\,2004dk and one $4$\,hour observation of SN\,2012ap. Daily observations of 3C286 were used for absolute flux calibration, and short interleaved observations of the phase calibrators (J1620+0036 for SN\,2004dk and J0501$-$0159 for SN\,2012ap) were used for phase calibrations. Images of the SNe fields were produced by the \texttt{CASA} task \texttt{clean} in an interactive mode. The image RMS was calculated using the \texttt{CASA} task \texttt{imstat}, and the sources at the phase centers were fitted with the \texttt{CASA} task \texttt{imfit}. We estimated the uncertainty of the peak flux density to be a quadratic sum of the error produced by \texttt{CASA} task \texttt{imfit} and a $5$ per cent calibration error.

We conducted eight observations of SN\,2012dy with ATCA for a total of $39$ hours -- with the project codes C3363 (PI Murphy) and C3442 (PI Horesh) -- in L/S-band ($1.1$--$3.1$\,GHz) and simultaneously in  C-band ($4.5$--$6.5$\,GHz) and X-band ($8.0$--$10.0$\,GHz).
For all observations we used the ATCA primary calibrator source PKS 1934$-$638 as the bandpass and flux calibrator, with the calibrator sources PKS 2117$-$642 and PKS 2204$-$540 for phase calibration scans in L-band and C/X-band respectively. We used the \texttt{Miriad} software \citep{1995ASPC...77..433S} to reduce and image the data, using \texttt{mfclean} with Briggs weighting and a \texttt{robust} parameter of $0.0$, with multi-frequency synthesis (\texttt{mfs}) deconvolution. We used the \texttt{Miriad} \texttt{imfit} routine to fit the point sources at the central frequencies for each full band ($2.1, 5.5, 9.0$\,GHz). We calculated the uncertainties as the quadrature sum of the fitted error, the image RMS, and a $10$ per cent flux scaling error.

\section{Methodology}
\label{sec: Methodology}

\subsection{Sample Selection}
\label{subsec: Sample Selection}

We compiled a list of known CCSNe -- as of October 2019 -- from a number of astronomical databases in order to obtain a set of candidate CCSNe with late-time radio emission. This cutoff date was chosen because ASKAP had completed three observing epochs of all VAST-P1 fields by the end of October 2019 (see Table 2 in \cite{2021PASA...38...54M}). After removing $534$ duplicate objects there were $10\,700$ CCSne including $3\,833$ from the Transient Name Server 
\citep[TNS\footnote{\url{https://www.wis-tns.org/}};][]{2021AAS...23742305G}, $3\,046$ from the Asiago supernova catalog \footnote{\url{https://heasarc.gsfc.nasa.gov/W3Browse/all/asiagosn.html}} \citep{1999A&AS..139..531B}, $500$ from Wiserep\footnote{\url{https://www.wiserep.org/}} \citep{2012PASP..124..668Y}, $900$ from SDSS-II Supernova Survey\footnote{\url{http://classic.sdss.org/supernova/snlist_confirmed.html}} \citep{2008AJ....135..338F}, and $2\,421$ from SIMBAD\footnote{\url{http://simbad.u-strasbg.fr/simbad}} \citep{2000A&AS..143....9W}.

From this collection of $10\,700$ we reduced the sample to $3\,658$ CCSNe by filtering out objects with declinations $>+40^{\circ}$, retaining only Type II and Type Ib/c SNe, and removing sources with $z>0.15$ or with unknown redshifts. This is because $z=0.15$ represents the limiting distance ($\sim700$\,Mpc) for ASKAP to make a $5\sigma$ detection of three times the maximum peak luminosity for CCSNe at $887.5$\,MHz: $L_{\nu}\sim10^{29}$\,erg s$^{-1}$Hz$^{-1}$ --- see Appendix \ref{subsec: Appendix A} for the sensitivity limit calculation. On average most CCSNe have a much lower peak luminosity around $L_{\nu}\sim 10^{25.5}$\,erg s$^{-1}$Hz$^{-1}$ \citep{2021ApJ...908...75B} which, with the same calculation, would correspond to a distance of $\sim8$\,Mpc ($z=0.002$). 
We cross-matched the sky coordinates of the $3\,658$ CCSNe with sources in the RACS-low catalogue using a $15$ arcsec match radius, finding matches for $29$ CCSNe --- see Appendix \ref{subsec: Appendix B} for the complete candidate list.

We generated lightcurves of radio detections and limits for these $29$ cross-matched CCSNe using data from VAST and RACS-low, as well as from archival surveys like the Faint Images of the Radio Sky at Twenty-Centimeters \citep[FIRST;][]{1995ApJ...450..559B} survey, the National Radio Astronomy Observatory (NRAO) VLA Sky Survey \citep[NVSS;][]{1998AJ....115.1693C}, and VLASS. The NVSS and FIRST surveys were carried out at $1.4$\,GHz and have typical RMS sensitivities of $0.45$\,mJy beam$^{-1}$ and $0.13$\,mJy beam$^{-1}$, respectively. These lightcurves, and the more comprehensive ones shown in Section \ref{sec: Results}, were generated relative to the optical discovery/explosion date found in the OpenSNe catalogue\footnote{\url{https://sne.space/}} \citep{2017ApJ...835...64G}.

We inspected these lightcurves, as well as radio image cutouts, and identified eight CCSNe with an ASKAP radio detection positionally offset from the nuclear region of the host galaxy: SN\,1996aq, SN\,2003bg, SN\,2004dk, SN\,2012ap, SN\,2012dy, SN\,2013bi, SN\,2016coi, SN\,2017gmr --- see Table \ref{tab:InterestingCandidates} for additional information. We selected these CCSNe as candidates for late-time re-brightening as the radio emission was detected $10^{2}$--$10^{3}$\,days after their initial optical discovery; In the cases of SN\,2012dy and SN\,2013bi the RACS-low observations were the first detections of radio emission from the sources. 

There were a further six of the $29$ cross-matched CCSNe which we also considered as potential candidates for late-time radio re-brightening: SN\,2002hy, SN\,2004gg, SN\,2006O, SN\,2008de, SN\,2008fi, SN\,2017hyh. We did not include these sources in our analysis because the ASKAP detection appears to be contaminated by radio emission from host galaxy. We list these sources and the full sample of cross-matched CCSNe in Appendix \ref{subsec: Appendix B}.

\begin{table*}
\caption{Primary CCSNe candidates for late-time radio re-brightening identified and analysed for this work. The type, redshift, discovery date, and distance are taken from the OpenSNe catalogue. The classification column refers to the category of re-brightening: a  re-brightening event due to interaction with a single CSM shell, variability due to interaction with CSM structure or other intrinsic processes, or refractive scintillation. We do not provide a classification for SN\,2013bi due to the lack of detections. The variability of SN\,2017gmr is consistent with scintillation but more data are required to conclusively classify the variability.}
\begin{tabular}{llllrccc}
\hline
 Name   &  Type   & RA       & Dec          & Redshift     &   Disc. [MJD]   &   Dist. [Mpc]  &  Classification \\
\hline
SN 1996aq       & SN Ic     & 14:22:22.7 & $-$00:23:23 & 0.0055  &            50312 & 23.9   &   Variability\\
SN 2003bg       & SN IIb    & 04:10:59.4 & $-$31:24:49  & 0.0045  &            52695 & 19.5   &   Re-brightening\\
SN 2004dk       & SN Ib     & 16:21:48.9 & $-$02:16:17  & 0.0052  &            53216 & 22.6   &   Re-brightening\\
SN 2012ap       & SN Ic-BL  & 05:00:13.7 & $-$03:20:50  & 0.012   &            55967 & 53.3   &   Variability\\
SN 2012dy       & SN II     & 21:18:50.7 & $-$57:38:42  & 0.0104  &            56142 & 45.3   &   Re-brightening\\
SN 2013bi      & SN IIp    & 18:25:02.1 & $+$27:31:53  & 0.016   &            56375 & 69.6   &   ...\\
SN 2016coi      & SN Ic-BL  & 21:59:04.1 & $+$18:11:11  & 0.0036  &            57535 & 15.6   &   Variability\\
SN 2017gmr      & SN II     & 02:35:30.2  & $-$09:21:14  & 0.005   &            58000 & 21.7    &   Scintillation\\
\hline
\end{tabular}

\label{tab:InterestingCandidates}
\end{table*}

\subsection{Modelling \& Analysis}
\label{subsec: Modelling & Analysis}

To analyse the appearance and evolution of the radio re-brightening events presented in this work, we adopted the standard model of synchrotron emission \citep{1982ApJ...259..302C} which describes the interaction between the propagating SN shockwave and the surrounding material. We assume a CSM density $\rho_{\rm CSM}\propto r^{-s}$ that is constant in time, unlike the outer density profile of the SN which has a power-law $\rho_{\rm SN}\propto r^{-n}t^{n-3}$. Following \cite{1982ApJ...259..302C} we require that $n>5$. 
From \cite{1998ApJ...499..810C} we adopt the expression for the radio flux density as a function of time and frequency:
\begin{multline}
    \label{eq: radio lightcurves}
    F_{\nu}(t)=1.582F_{\nu_{\rm c}}(t_{\rm c})\left(\frac{t}{t_{\rm c}}\right)^a\left(\frac{\nu}{\nu_{\rm c}}\right)^{5/2}\\\times\left(1-\exp\left[-\left(\frac{t}{t_{\rm c}}\right)^{-(a+b)}\left(\frac{\nu}{\nu_{\rm c}}\right)^{-(\gamma+4)/2}\right] \right) 
\end{multline}

\noindent
where $F_{\nu}\propto t^a$ is the power law in the optically-thick regime; $F_{\nu}\propto t^{-b}$ is the power law in the optically-thin regime; and $t_{\rm c}$, $\nu_{\rm c}$ are the peak time and frequency at which the emission transitions from optically-thick to  optically-thin. 

When the spectral peak is given, \cite{1998ApJ...499..810C} provides the following equations to calculate the radius of a fast-moving shock $R_{\rm p}$ at the leading edge of the emitting region, as well its magnetic field strength $B_{\rm p}$:

\begin{align}
    \label{eq: shock radius}
     R_{\rm p} = \, & \left[ \frac{6{c_{6}}^{\gamma+6}F_{\rm p}^{\gamma+6}D^{2\gamma+12}}{\alpha_{\epsilon} f(\gamma-2)\pi^{\gamma+5}c_5^{\gamma+6}E_{\rm l}^{\gamma-2}} \right]^{1/(2\gamma+13)}\left(\frac{\nu}{2c_1} \right)^{-1}
\end{align}

\begin{align}
    \label{eq: magnetic field}
     B_{\rm p} = \, & \left[ \frac{36\pi^3c_{5}}{\alpha_{\epsilon}^2 f^2(\gamma-2)^2c_6^{3}E_{\rm l}^{2(\gamma-2)}F_{\rm p}D^2} \right]^{2/(2\gamma+13)}\left(\frac{\nu}{2c_1} \right)
\end{align}

\noindent
where $\alpha_{\epsilon}$ is the ratio of electron and magnetic energy densities, $D$ is the distance to the SN, $F_{\rm p}$ is the peak flux density, $\nu$ is the frequency, $f$ is the geometric filling factor for the emission, $E_{\rm l}=0.51$\,MeV is the electron rest energy, and $c_1,c_5,c_6$ are constants (See Appendix \ref{subsec: Appendix D}).

In this work we assume an equipartition of electron and magnetic energies $\epsilon_{\rm e} = \epsilon_{\rm B}=0.1$ $\rightarrow$ $\alpha_{\epsilon}=1$ --- although there have been a number of recent examples of deviation from equipartition \citep[e.g.][]{2020ApJ...903..132H,2022A&A...666A..82R}. We also assume an emission filling factor $f=0.5$, and an electron energy spectral index $\gamma=3$.

The shock radius can be used to calculate the mean shock velocity as $v_{\rm sh}=R_{\rm p}/t$. A typical assumption is that the magnetic energy density, $B^2/8 \pi$, is a fraction, $\epsilon_{\rm B}$, of the post shock energy density, $\frac{9}{8}\rho_{\rm CSM} v_{\rm sh}^2$. In the case of a constant mass-loss in a steady wind $v_{\rm w}$, as is often assumed in the CSM interaction model, we can define CSM density simply as: 

\begin{equation}
    \label{eq: csm density}
    \rho_{\rm CSM} = \frac{\dot{M}}{4\pi r^2 v_{\rm w}}.
\end{equation}
This results in the mass-loss rate:

\begin{multline}
    \label{eq: mass-loss}
    \dot{M} =  5.2 \times 10^{-7} \, \left( \frac{\epsilon_{\rm B}}{0.1} \right)^{-1} \left( \frac{B}{1 \, \rm{G}} \right)^{2} \\\times \left( \frac{t}{10 \, \rm{days}} \right)^{2} \left( \frac{v_{\rm w}}{100 \, \rm{km~s^{-1}}} \right) \, \rm M_{\odot}~yr^{-1}
\end{multline}

\noindent
which we can calculate for a range of wind velocities $v_{\rm w}$ as part of our general investigation into the physical properties of the circumstellar region. For our calculations we use the wind velocities $10\,\rm{km~s^{-1}}$ and $1\,000\,\rm{km~s^{-1}}$ for Type II and Type Ib/c SNe respectively \citep{2014ARA&A..52..487S}.

We use the \texttt{emcee} Python package \citep{2013PASP..125..306F} to implement a Markov chain Monte Carlo (MCMC) sampler and fit the observations to the \cite{1998ApJ...499..810C} model. We used $100$ walkers and a burn-in period of $1000$ iterations. For our model we used the more general form of Equation \ref{eq: radio lightcurves} to include a full parameterisation of the multi-frequency data 
\begin{multline}
    \label{eq: parameterised model}
    F_{\nu}(t)=1.582F_{\nu_{\rm c}}(t_{\rm c})\left(\frac{t}{t_{\rm c}}\right)^a\left(\frac{\nu}{\nu_{\rm c}}\right)^{\alpha}\\\times\left(1-\exp\left[-\left(\frac{t}{t_{\rm c}}\right)^{-(a+b)}\left(\frac{\nu}{\nu_{\rm c}}\right)^{-(\alpha+\beta)}\right] \right) 
\end{multline}
assuming a peak frequency of $887.5$\,MHz. 

This model is applied to SNe which produced detectable radio emission at early times, using a $\chi^2$ log likelihood function
\begin{equation}
    \label{eq: likelihood function}
    \log\left[\mathcal{L}(F|\theta, t,\delta F)\right]= -\frac{1}{2}\sum\left(\frac{F-F_{\nu}(t,\theta)}{\sqrt{\delta F}}\right)^2 
\end{equation}

\noindent
where $F$ is the flux density measurement, $\delta F$ the flux density errors, and $\theta$ the parameters $F_{\rm p},t_{\rm p},\alpha,\beta,a$, and $b$. We use flat prior distributions for these parameters, over the following prior ranges: $10^{-2} < F_{\rm p} < 10^2$; $10^{-1} < t_{\rm p} < 2\times10^4$; $-5.0 < \alpha < 5.0$; $-5.0 < \beta < 5.0$; $-5.0 < a < 5.0$; $-5.0 < b < 5.0$. For further details about the MCMC algorithm, see \cite{2013PASP..125..306F}.

We fit the multi-frequency data from $\Delta t < 3\times10^3$\,days. We do not fit this model to the SNe with few or no radio detections within the first $\sim$10$^2$\,days after optical discovery.

For all SNe we also extract astrophysical parameters of the shock and surrounding medium, making the simplifying assumption that the peak in the lightcurve follows the transition of the shock from an optically-thick to an optically-thin region --- using Equations \ref{eq: shock radius}, \ref{eq: magnetic field}, \ref{eq: mass-loss}. We discuss the validity of this assumption in Section \ref{sec: Discussion}.

We also quantify the variability of the emission using the modulation index 
\begin{equation}
    \label{eq:modulation index}
    V=\frac{1}{{\langle F\rangle}}\sqrt{\frac{N}{N-1}\langle{F^2\rangle}-\langle{F\rangle}^2}
\end{equation}
and the reduced $\chi^2$ parameter (relative to a constant model)
\begin{equation}
    \label{eq: reduced chi-squared}    
    \eta=\frac{N}{N-1}\left(\langle w F^2 \rangle \frac{\langle w F \rangle ^2}{\langle w\rangle}\right)
\end{equation}

\noindent
where $N$ is the number of measurements and $w_i=1/\sigma_i^{2}$ is the measurement weight defined with $\sigma_i$ being the uncertainty corresponding to the $i$-th measurement. We use $\langle  \rangle$ to denote the arithmetic mean of a value across the  measurements.
The $V,\eta$ metrics provide measures of the degree and significance of the variability, respectively.

For comparison we also calculate the expected variability index $m$ caused by refractive interstellar scintillation (RISS) at $887.5$\,MHz, assuming a scattering screen at a distance of $1$\,Mpc. For this we use the \texttt{RISS19} package \citep{2019arXiv190708395H} 
For observations at $887.5$\,MHz -- or generally $\leq 1$\,GHz -- extragalactic sources like SNe are in the strong scattering regime. In this regime RISS produces lower variation in time compared to diffractive scintillation, which is uncommon in extragalactic radio sources \citep{2019arXiv190708395H}.

\section{Results}
\label{sec: Results}

In this section we describe the results from our review of the CCSNe that were found to be sources potentially undergoing periods of late-time radio re-brightening. Besides the databases (see Section\,\ref{subsec: Sample Selection}), our archival review included a thorough search for reported emission at optical, radio, or X-ray wavelengths in the literature --  Astrophysics Data System (ADS\footnote{\url{https://ui.adsabs.harvard.edu/}}) and Astronomer's Telegram (ATel\footnote{\url{https://www.astronomerstelegram.org/}}) -- as well as in the archives of NASA/IPAC Extragalactic Database (NED\footnote{\url{https://ned.ipac.caltech.edu/}}), High Energy Astrophysics Science Archive Research Center (HEASARC\footnote{\url{https://heasarc.gsfc.nasa.gov/}}), Panoramic Survey Telescope And Rapid Response System (PanSTARRS\footnote{\url{https://ps1images.stsci.edu/cgi-bin/ps1cutouts}}), VLASS\footnote{\url{http://cutouts.cirada.ca/}}, CSIRO ASKAP Science Data Archive (CASDA\footnote{\url{https://data.csiro.au/collections/domain/casdaObservation/search/}}), and Australia Telescope Online Archive (ATOA\footnote{\url{https://atoa.atnf.csiro.au/query.jsp}}). 
One motivation for this search was to demonstrate the role of ASKAP to commensally study the late-time evolution of CCSNe and  detect new CCSNe re-brightening events.

\subsection{SN 1996aq}
\label{subsec: SN 1996aq}
SN\,1996aq was discovered and classified as a SN Ic by \cite{1996IAUC.6454....1N} but was not detected in radio until 2009. It was observed in the C- and X- bands ($4.9$\,GHz, $8.5$\,GHz) by \cite{2009CBET.1714....1S} with the VLA and with the Giant Metrewave Radio Telescope (GMRT) at $617$\,MHz \citep{2009ATel.1974....1C}. More recently SN\,1996aq has been detected in the L/S-bands with the VLA (Project Code:16A-210), VLASS \citep{2020RNAAS...4..175G}, RACS-low, and multiple epochs of VAST-P1.

\begin{figure}
\centering
\includegraphics[width=8.5cm]{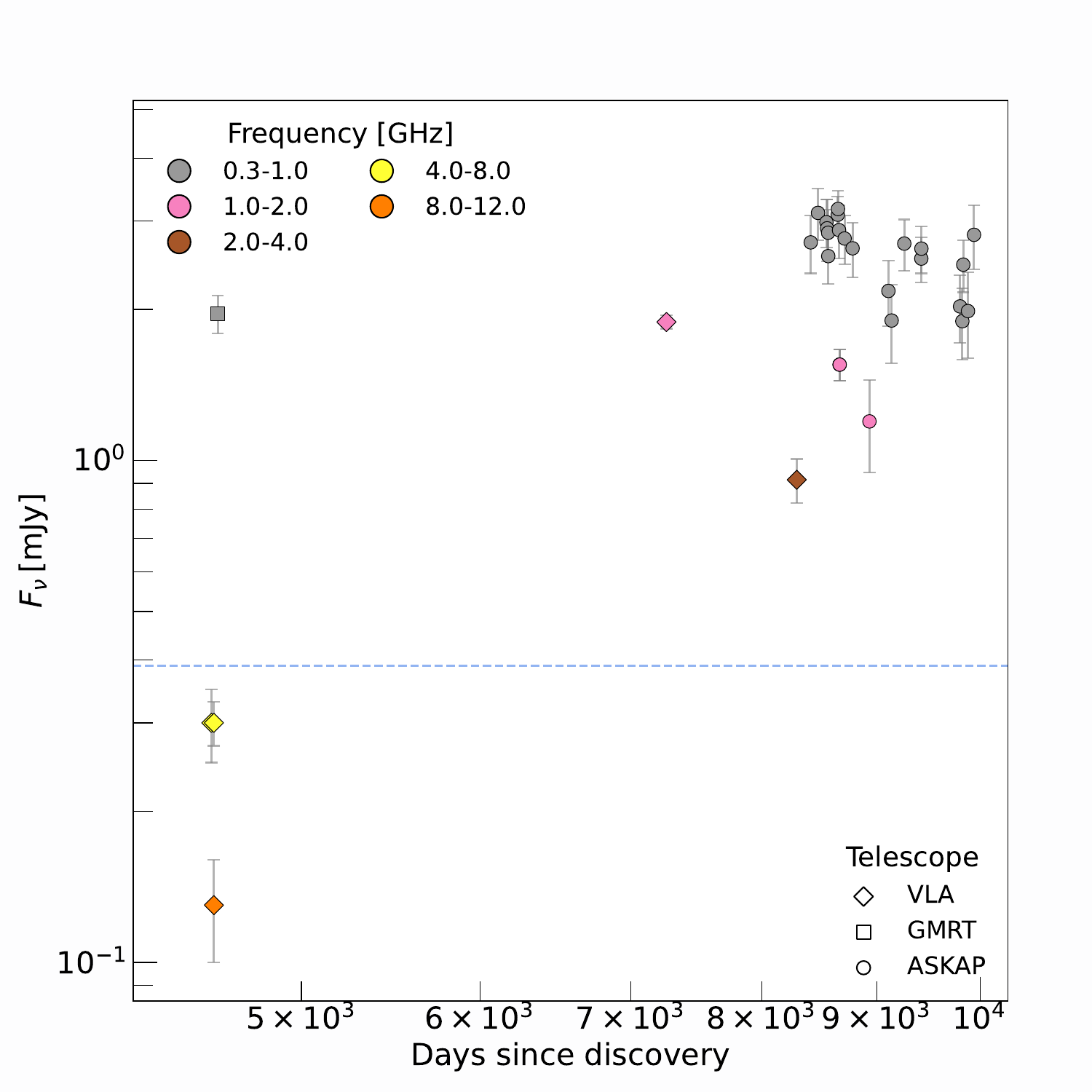}
\caption{Radio detections of SN\,1996aq from the literature and CASDA, including all public ASKAP observations as of December 2023, binned into frequency bins and coloured accordingly. We use diamond markers for VLA detections, squares for GMRT, and circles for ASKAP. The dashed blue line represents the FIRST $3\sigma$ limit for any potential confusing host emission.}
\label{fig: SN 1996aq lightcurve}
\end{figure}

We searched archival NVSS and FIRST observations for host emission or other contaminating radio sources near the SN coordinates. These observations were taken at $\Delta t=-538$\,days and $\Delta t=+720$\,days, respectively relative to the optical discovery, and were both observed at a central frequency of $1.4$\,GHz. There is no emission detected within $5$ arcsec of SN\,1996aq in either observation, with a $5\sigma$ limit $0.735$ mJy in the case of FIRST. By contrast the SN was detected with the VLA at $1.4$\,GHz with a flux density of $1.89\pm0.06$ mJy on $\Delta t = +7\,258$\,days. This VLA detection, offset by $<1$ arcsec from the SN coordinates, confirms that it is indeed SN\,1996aq, and not one of the nearby radio sources, that is producing this increased radio flux density.

We used the equations given in Section \ref{subsec: Modelling & Analysis} with the peak ASKAP flux density measurement of $F_{\rm p}=3.17\pm0.28$\,mJy beam$^{-1}$ at $855.5$\,MHz from $\Delta t=8\,648$\,days. We calculate a lower limit on the shock radius to be $R_{\rm p} \gtrsim 5.4\times10^{16}$\,cm and an upper limit on the magnetic field strength of $B_{\rm p}\lesssim0.097$\,G --- using Equations \ref{eq: shock radius}, \ref{eq: magnetic field}å.

We list the calculated values for SN\,1996aq, and for the other seven CCSNe in our sample, in Table \ref{tab: MaxValues}. Assuming that the emission is optically thin, we use Equation \ref{eq: mass-loss} to calculate the mass-loss rate. Assuming a wind velocity of $v_{\rm w }=1\,000\,\rm{km~s^{-1}}$ (see Section \ref{subsec: Modelling & Analysis}) we find that it would require a CSM density of $\rho_{\rm{CSM}}\approx6.2\times10^{-19}$\,g cm$^{-3}$ and a mean shockwave velocity $v_{\rm sh}\approx730$ km s$^{-1}$. We use the shock radius $R_{\rm p}$ with the time $\Delta t$ to calculate this mean shockwave velocity --- see Table \ref{tab: MaxValues} for values.
Since the \cite{1998ApJ...499..810C} equations are only valid in the high-velocity regime, there is some theoretical uncertainty in the values obtained for $B_{\rm p}$ and $R_{\rm p}$. There are additional theoretical uncertainties in the calculated values from the assumptions of our equipartition model -- see Section \ref{subsec: Modelling & Analysis} -- as well as the measurement uncertainties.

By comparing flux densities from RACS-low and VLASS -- extracting a flux density with the \texttt{CASA} \texttt{imfit} task -- we obtained a spectral index of $\alpha=-0.91$. From Figure \ref{fig: SN 1996aq lightcurve} we can see that SN\,1996aq is in the midst of a clear period of late-time re-brightening. We found that the variability of this re-brightening over time is not consistent with a power-law.

\begin{table*}
    \centering
    \caption{The brightest radio detections of each SN in our sample. We use the peak flux densities and observing frequencies to obtain limits on the radius of the emission region, as well as the magnetic field strength, CSM density, number density, and shock velocity. These peak flux densities are a limit on the spectral peak, meaning that the true peak flux density should be higher and the true peak frequency lower. The brightest measurements for five of the eight SNe in this sample come from late-time ($\Delta t>10^3$\,days) emission, four of which were detected with ASKAP.
    Detections at $0.887$\,GHz and $0.855$\,GHz are from ASKAP with the other detections observed with the VLA.}
    \setlength{\tabcolsep}{4pt}
\begin{tabular}{llllcccccl}
\hline
 Name   &  $\Delta t$ [days]  & $F_{\nu}$ [mJy beam$^{-1}$] & $\nu$ [GHz] & $L_{\nu}$ [erg s$^{-1}$ Hz$^{-1}$] 
 & $R_{\rm p}$ [cm] & $B_{\rm p}$ [G] & $\rho_{\rm{CSM}}$ [g cm$^{-3}$]  & $n_{\rm p}$ [cm$^{-3}$] & $v_{\rm sh}$ [km s$^{-1}$] \\
\hline

SN 1996aq & $8\,648$ & $3.17 \pm 0.28$ & $0.855$ & $ {\sim}1.4\times 10^{27}$ & 
$ {\gtrsim}5.4\times10^{16}$ & $ {\lesssim}0.097$ & $\approx6.2\times10^{-19}$ & $\approx3.7\times10^5$ & $\approx730$\\

SN 2003bg &  {$23$} &  {$106.3 \pm 2.2$} &  {$22.5$} &  {$\sim5.2\times 10^{28}$} & 
 {$\gtrsim1.2\times10^{16}$} &  {$\lesssim 1.7$} &  {$\approx3.1\times10^{-20}$} &  {$\approx1.9\times10^3$} &  {$\approx58\,000$}\\

SN 2004dk & $5\,994$ & $21.53 \pm 4.31$ & $0.34$ & $ {\sim}4.5\times 10^{26}$ & 
$ {\gtrsim}8.1\times10^{16}$ & $ {\lesssim}0.043$ & $\approx2.7\times10^{-20}$ & $\approx1.6\times10^4$ & $\approx1\,500$\\

SN 2012ap & $4\,279$ & $10.98 \pm 1.08$ & $0.887$ & $ {\sim}3.9\times 10^{28}$ & 
$ {\gtrsim}2.6\times10^{17}$ & $ {\lesssim}0.071$ & $\approx3.6\times10^{-21}$ & $\approx2.1\times10^3$ & $\approx7\,000$\\

SN 2012dy & $2\,580$ & $18.37 \pm 1.12$ & $0.887$ & $ {\sim}4.6\times 10^{28}$ & 
$ {\gtrsim}2.8\times10^{17}$ & $ {\lesssim}0.069$ & $\approx1.1\times10^{-21}$ & $\approx6.5\times10^2$ & $\approx13\,000$\\

SN 2013bi & $2\,595$ & $4.73 \pm 0.41$ & $0.887$ & $ {\sim}2.9\times 10^{28}$ & 
$ {\gtrsim}2.2\times10^{17}$ & $ {\lesssim}0.073$ & $\approx1.9\times10^{-21}$ & $\approx1.1\times10^3$ & $\approx10\,000$\\

SN 2016coi & $21$ & $24.0 \pm 1.0$ & $21.85$ & $ {\sim}7.5\times 10^{27}$ & 
$ {\gtrsim}4.8\times10^{15}$ & $ {\lesssim}2.08$ & $\approx2.2\times10^{-19}$ & $\approx1.3\times10^5$ & $\approx26\,000$\\

SN 2017gmr & $858$ & $1.89 \pm 0.39$ & $0.887$ & $ {\sim}1.1\times 10^{27}$ & 
$ {\gtrsim}4.8\times10^{16}$ & $ {\lesssim}0.10$ & $\approx8.8\times10^{-21}$ & $\approx5.3\times10^3$ & $\approx6\,500$\\
\hline
\end{tabular}

    \label{tab: MaxValues}
\end{table*}

\subsection{SN 2003bg}
\label{subsec: SN 2003bg}

SN\,2003bg was discovered by \cite{2003IAUC.8082....1W} and within a year its spectral features had evolved, resulting in a reclassification from a SN Ic to a SN IIb hypernova \citep{2009ApJ...703.1612H,2009ApJ...703.1624M}. Based on variability in the radio lightcurve, it was suggested that this transition was due to CSM interaction \citep{2006ApJ...651.1005S}.

\cite{2013A&A...557L...2M} suggested that the radio emission from SN\,2003bg may not be undergoing re-brightening but rather may be the modulated tail-end from the prompt emission decaying to lower frequencies --- noting its similarity to SN\,2001ig \citep{Ryder2004}. Similar comparisons were drawn to SN\,2011ei by \cite{2013ApJ...767...71M}, who suggested a two-component emission model to explain the deviation from an expected power-law decay in flux.

\begin{figure}
\centering
\includegraphics[width=8.5cm]{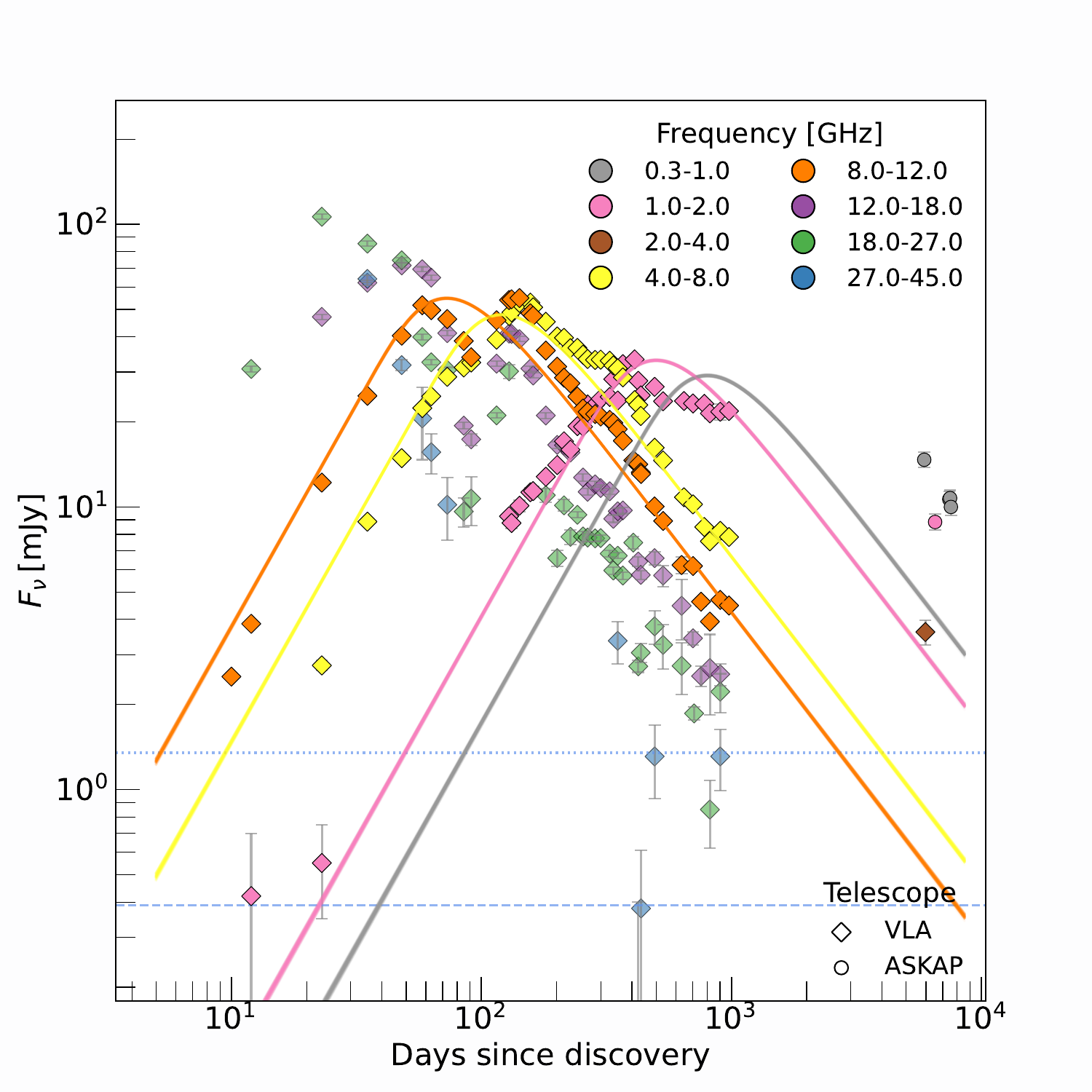}
\caption{We show the archival radio detections of SN\,2003bg \citep{2006ApJ...651.1005S, 2020RNAAS...4..175G} along with the modelled lightcurves obtained from fitting of the early-time ($\Delta t<10^3$\,days) data. We also plot detections from  all public ASKAP observations as of December 2023. The MCMC parameterised lightcurves are extrapolated from the multi-frequency data and plotted for $0.8875, 1.4, 5.5, 9.0$\,GHz. The data binned into frequency bins and coloured accordingly.  We use diamond markers for VLA detections and circles for ASKAP. The dashed and dotted blue lines represents the FIRST and NVSS $3\sigma$ limits, respectively,  for any potential confusing host emission.}
\label{fig: SN 2003bg lightcurve}
\end{figure}

The early-time data for SN\,2003bg fit the \cite{1998ApJ...499..810C} model -- see Figure \ref{fig: SN 2003bg lightcurve} -- including a possible break in the decline shortly after the spectral turnover. We provide the fitted parameters in Table \ref{tab:FitParameters} and the MCMC posterior distributions corner plots in  Appendix \ref{subsec: Appendix F}. Based on our modelling  it appears that the observed RACS emission is $\sim 5\times$ brighter than would be expected from the tail of the prompt emission. The same is true for the available VLASS measurement of SN\,2003bg, which is a factor of $\sim 3\times$ brighter than the extrapolated value. 

While a two-component model may provide a good fit with the early-time data -- see \cite{2006ApJ...651.1005S} -- it does not explain the significant late-time re-brightening. Further observation and modelling of this SN may provide insights into  density enhancements in the circumstellar region following pre-eruption mass-loss episodes.

\subsection{SN 2004dk}
\label{subsec: SN 2004dk}

SN\,2004dk was one of the better-studied CCSNe in our sample, with multiple VLA observations of the region from the periods before and shortly after its discovery by \cite{2005PASP..117..132R}, who classified it as a SN Ib.
\cite{2012ApJ...752...17W} were the first to identify the re-brightening of SN\,2004dk and subsequent spectral analyses have noted the initial interaction with the hydrogen-rich CSM occurred $\Delta t\sim 1\,660$\,days after discovery. Optical spectroscopy conducted by \cite{2018MNRAS.478.5050M} at $4\,684$ days confirmed that the interaction with the H-rich CSM was ongoing.

\begin{figure}
\centering
\includegraphics[width=8.5cm]{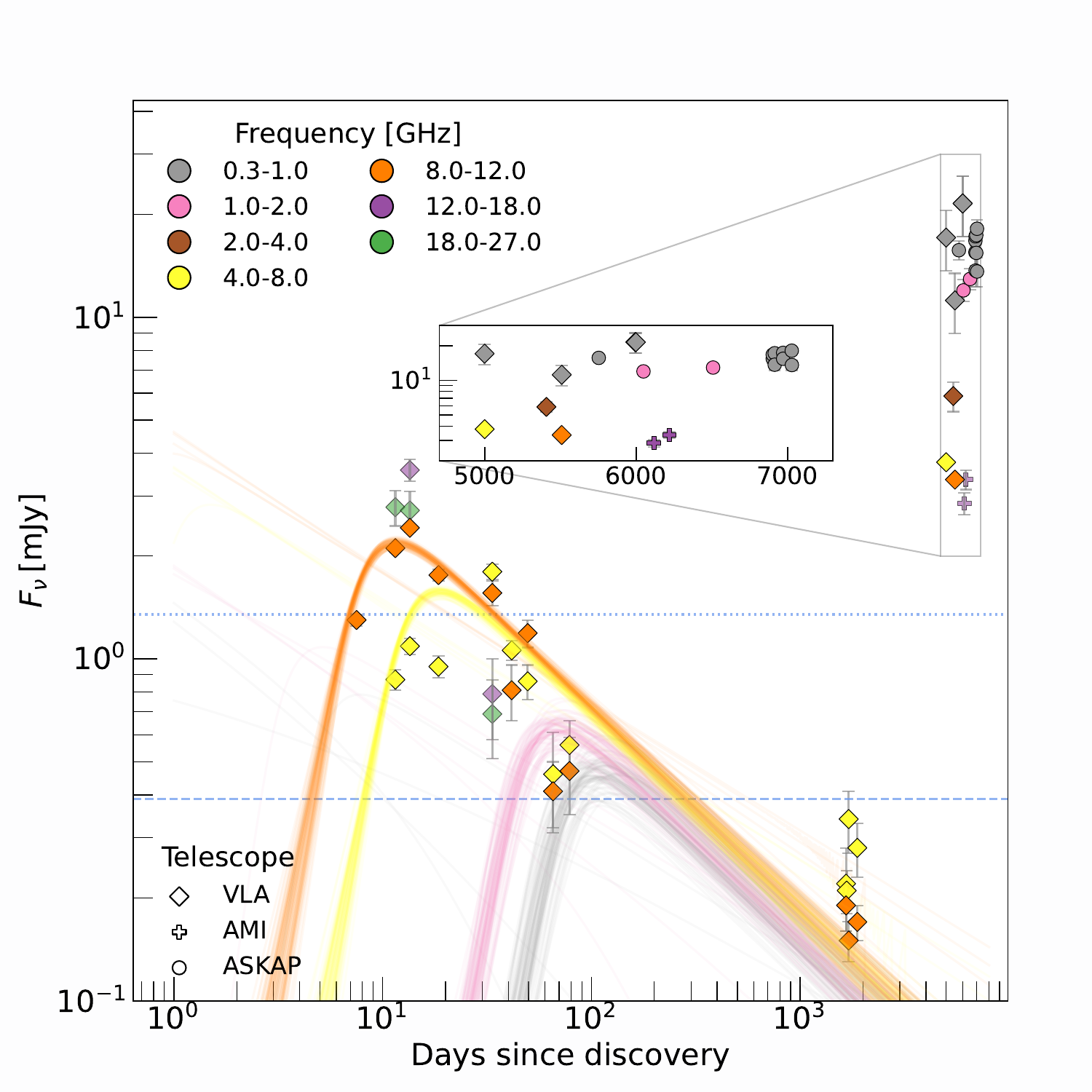}
\caption{We show the archival radio detections of SN\,2004dk \citep{2012ApJ...752...17W,2021ApJ...923...32B} along the modelled lightcurves obtained from fitting of the early-time ($\Delta t<3\times10^3$\,days) data. We also plot detections from  all public ASKAP observations as of December 2023. The MCMC parameterised lightcurves are extrapolated from the multi-frequency data and plotted for $0.8875, 1.4, 5.5, 9.0$\,GHz, with the data binned into frequency bins and coloured accordingly. The inset shows a zoomed version of the late-time ($\Delta t>3\times10^3$\,days) data.  We use diamond markers for VLA detections, pluses for AMI-LA, and circles for ASKAP. The dashed and dotted blue lines represents the FIRST and NVSS $3\sigma$ limits, respectively,  for any potential confusing host emission.}
\label{fig: SN 2004dk lightcurve}
\end{figure}

\cite{2021ApJ...923...32B} also reported the radio re-brightening of  SN\,2004dk at low frequencies ($340$\,MHz) with the VLA Low-band Ionosphere and Transient Experiment \citep[VLITE;][]{2016SPIE.9906E..5BC} --- around the same time that RACS-low detected the SN.
We conducted follow-up observations with AMI-LA and detected $2.85\pm0.21$\,mJy beam$^{-1}$ peak flux density and $4.67\pm0.46$\,mJy integrated flux density at $15.5$\,GHz. This emission is $\sim3\times$ brighter than the $15$\,GHz VLA observations from $\Delta t=33$ days \citep{2012ApJ...752...17W}. We also see an order of magnitude increase in the VLA C- and X-band emission after $\Delta t \sim 10^3$\,days, as well as a similar increase in the $1$--$2$\,GHz emission at late times compared to the NVSS $3\sigma$ limit --- see Figure \ref{fig: SN 2004dk lightcurve}.

The lightcurve evolution and re-brightening of SN\,2004dk presented in Figure \ref{fig: SN 2004dk lightcurve} appears similar to that of SN\,2014C, which has been modelled as the late-time interaction of the SN shock with a dense circumstellar shell comprised of material from a previously stripped envelope \citep{2017MNRAS.466.3648A}.

\begin{table*}
    \centering \caption{\protect\cite{1998ApJ...499..810C} model parameters for early-time emission. For each of the fitted CCSNe we give the posteriors estimated with MCMC. We fit the early-time data -- $\Delta t <10^3$\,days for SN\, 2003bg and $\Delta t <3\times10^3$\,days for SN\, 2004dk and SN\,2016coi -- } with the free parameters $\alpha$, $\beta$ --- see Equation \ref{eq: parameterised model}. These are the exponents for frequency dependent terms with $\alpha$ and $\alpha + \beta$, in place of $5/2$ and $(\gamma + 4)/2$ in Equation \ref{eq: radio lightcurves}. This $\alpha$ is unrelated to the $\alpha_{\epsilon}$ used to denote the energy density ratio in Equations \ref{eq: shock radius},\ref{eq: magnetic field}. The archival flux density measurement uncertainties used in the MCMC fit are taken from the literature --- see Section \ref{sec: Observations} for the calibration uncertainties included for data from ATCA, AMI-LA, and ASKAP. See Appendix \ref{subsec: Appendix F} for corresponding posterior distribution corner plots.
    \begin{tabular}{llllrrr}
\hline
 Name           & $F_{\rm p}$   & $t_{\rm p}$       & $\alpha$          & $\beta$     &   $a$   &   $b$\\
\hline
SN 2003bg        & {$29.16^{+0.14}_{-0.14}$}    & {$804.75^{+5.04}_{-4.96}$} & {$1.91^{+0.01}_{-0.01}$}  & {$0.92^{+0.004}_{-0.004}$}  &  {$1.58^{+0.01}_{-0.01}$} & {$1.15^{+0.004}_{-0.004}$}\\

SN 2004dk       & {$0.40^{+0.04}_{-0.03}$}    & {$80.16^{+10.93}_{-10.08}$} & {$3.82^{+0.38}_{-0.31}$}  & {$-0.11^{+0.05}_{-0.06}$}  &  {$3.25^{+0.43}_{-0.36}$} & {$0.58^{+0.02}_{-0.02}$}\\

SN 2016coi       & $11.02^{+0.37}_{-0.36}$    & $456.48^{+15.73}_{-15.55}$ & $2.13^{+0.05}_{-0.05}$  & $0.87^{+0.03}_{-0.03}$  &  $1.81^{+0.04}_{-0.04}$ & $1.12^{+0.03}_{-0.03}$\\
\hline
\end{tabular}

    \label{tab:FitParameters}
\end{table*}

\subsection{SN 2012ap}
\label{subsec: SN 2012ap}

Following the discovery of SN\,2012ap \citep{2012CBET.3037....1J} a number of groups reported on the optical spectroscopy and photometry in the first $100$--$272$\,days, and its classification as a SN Ic-BL \citep{2015ApJ...799...51M,2016MNRAS.458.2973P}.  

Early-time VLA detections showed the peak frequency shift from $11.82\pm0.48$\,GHz to $4.54\pm0.10$\,GHz within $\Delta t=38$\,days of discovery \citep[see Table 1,][]{2015ApJ...805..187C}. Over that same period, the corresponding peak flux densities declined from $5.85\pm0.58$\,mJy to $4.20\pm0.10$\,mJy. This decay can be explained by the standard SSA emission mechanism \citep{1998ApJ...499..810C} but \cite{2015ApJ...805..187C} noted that the evolution occurred much faster than other SNe Ibc. They calculated a lower limit of $0.4c$ on the initial expansion velocity, suggesting that there may have been an associated Gamma-Ray Burst (GRB), though none was found.

\begin{figure}
\centering
\includegraphics[width=8.5cm]{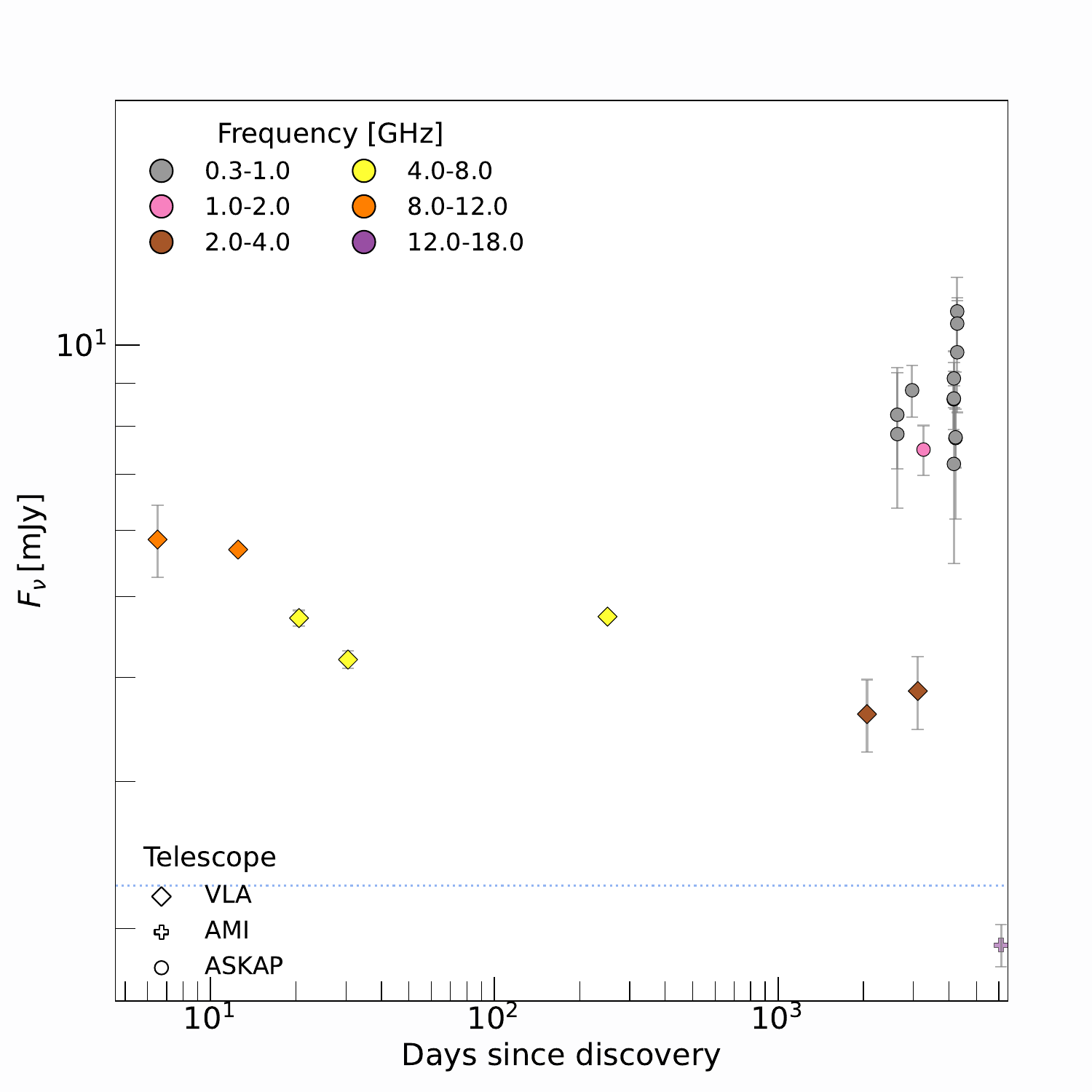}
\caption{Archival radio detections of SN\,2012ap from the literature \citep{2015ApJ...805..187C,2020RNAAS...4..175G} and CASDA, including all public ASKAP observations as of December 2023, binned into frequency bins and coloured accordingly. We use diamond markers for VLA detections, pluses for AMI-LA, and circles for ASKAP. The dotted blue lines represents the NVSS $5\sigma$ limit for any potential confusing host emission.}
\label{fig: SN 2012ap lightcurve}
\end{figure}

We have since seen with VLASS \citep{2020RNAAS...4..175G} and RACS-low that the peak frequency has continued to evolve more slowly at later times, as seen in Figure \ref{fig: SN 2012ap lightcurve}. This is reasonable considering the deceleration of the shock that would occur while sweeping up additional mass as it propagates through the CSM and into the ISM.

Following early VLA and GMRT observations, \cite{2015ApJ...805..187C} noted that SN\,2012ap represented the first case of a SN with rapidly decelerating mildly relativistic ejecta but without a coincident GRB; there has only been one other SN with relativistic ejecta without an associated GRB \citep{2010Natur.463..513S}.

Consequently a VLASS study \citep{2021ApJ...923L..24S} -- which also included the sources  SN\,2003bg, SN\,2004dk, and SN\,2016coi in their sample --- suggested that, aside from the CSM interaction emission mechanism, the late-time radio re-brightening seen in SN\,2012ap could be generated by an off-axis jet and cocoon; we note that this would be the first evidence of such a system from a massive stellar explosion. 

\cite{2021ApJ...923L..24S} reported a clear VLASS detection coincident with the optical SN position -- with no host emission -- and our follow-up observation with AMI-LA detected a clear source of emission uncontaminated by emission from nearby sources. We used archival NVSS data for reference.

\subsection{SN 2012dy}
\label{subsec: SN 2012dy}

After its initial discovery by a member of the Backyard Observatory Supernova Search (BOSS) team \citep{2012CBET.3197....1B}, SN\,2012dy was classified as a SN IIb by the Public European Southern Observatory Spectroscopic Survey of Transient Objects \citep[PESSTO;][]{2015A&A...579A..40S}. PESSTO spectra taken at weekly intervals over $\sim$140\,days after discovery show the development of broad [O {\sc i}], Si\,{\sc ii}, Ca\,{\sc ii}, and [Ca\,{\sc ii}] emission features, which is indicative of CSM interaction.
Following this there were no observations of the object nor its host galaxy at any wavelength until $\sim8$ years later when we first detected SN\,2012dy with ASKAP.

\begin{table}
    \centering 
    \caption{We fit a power-law $F(\nu)=F_0(\nu/\nu_0)^{\alpha_{\rm fit}}$ to the SN\, 2012dy ATCA observations centred at $2.1, 5.5, 9.0$\,GHz to extrapolate the expected flux density $F_{\rm ex}$ at $0.8875$\,GHz. The observation on $\Delta t=3\,422$ only includes the bands centred at $2.1, 5.5$\,GHz.}
    \begin{tabular}{ccccc}
\hline
  $\Delta t$ [days]  & $\alpha_{\rm fit}$ &  $\nu_0$ [GHz] & $F_{0}$ [mJy] &  $F_{\rm ex}$ [mJy]\\
\hline

3\,220 & $-0.87$& $1.49$& $11.66$& $18.30$\\
3\,257 & $-0.81$& $1.49$& $11.11$& $16.90$\\
3\,350 & $-0.79$& $1.47$& $10.52$& $15.71$\\
3\,388 & $-0.84$& $1.50$& $11.05$& $17.13$\\
3\,422 & $-0.97$& $1.46$& $10.20$& $16.54$\\
3\,445 & $-1.00$& $1.26$& $10.10$& $14.38$\\
3\,472 & $-0.96$& $1.54$& $10.39$& $17.71$\\
3\,500 & $-0.93$& $1.55$& $10.34$& $17.32$\\

\hline
\end{tabular}

    \label{tab:SED_SN2012dy}
\end{table}

\begin{figure}
\centering
\includegraphics[width=8.5cm]{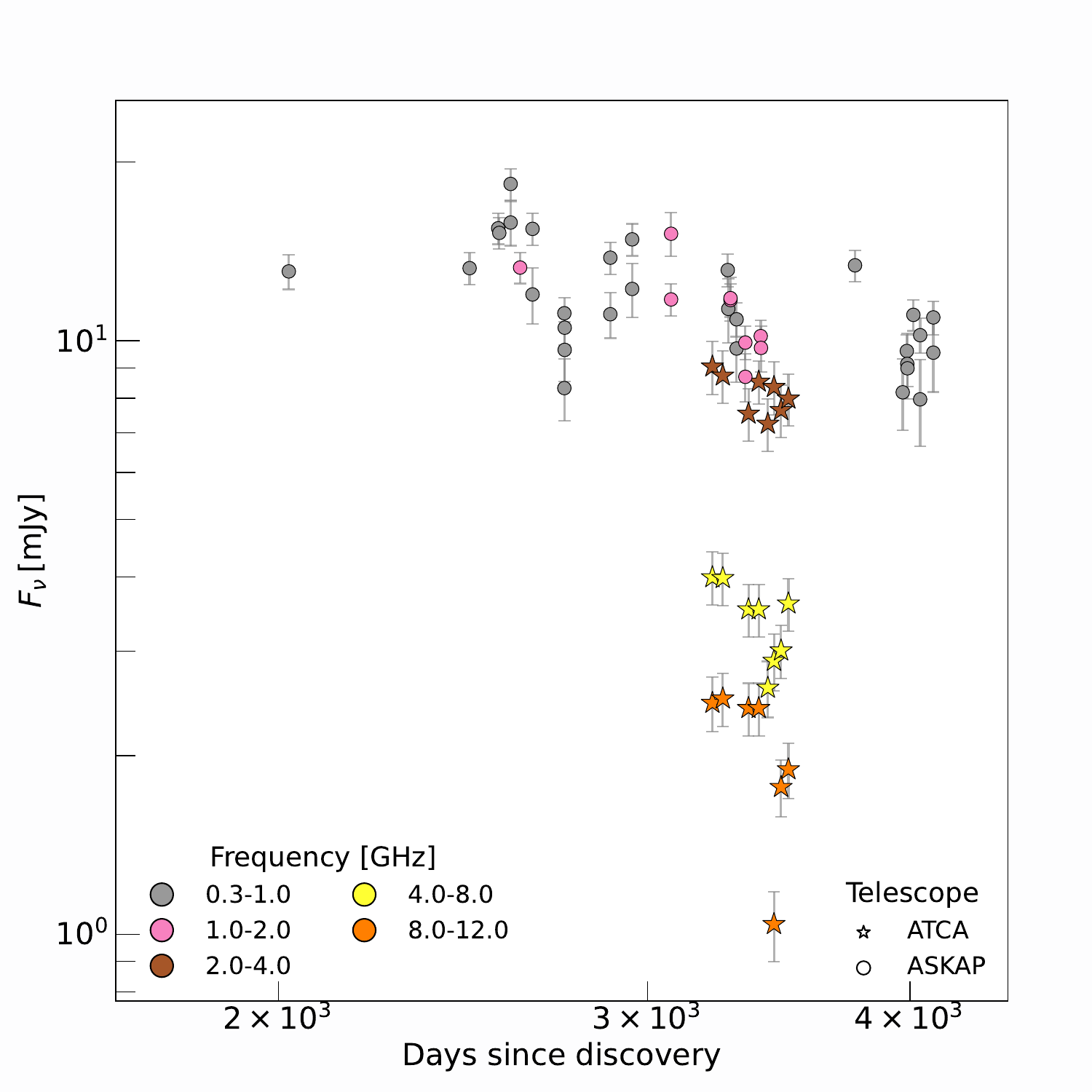}
\caption{Radio detections of SN\,2012dy from our ATCA observations and all public ASKAP data as of December 2023, binned into frequency bins and coloured accordingly. We use circle markers for ASKAP detections and stars for ATCA.}
\label{fig: SN 2012dy lightcurve}
\end{figure}

Given the flux density variability of the source observed across RACS-low and several VAST-P1 epochs -- see in Figure \ref{fig: SN 2012dy lightcurve} where RACS-low is the first radio detection of this SN -- we conducted follow-up observations with ATCA under an existing VAST NAPA proposal (C3363). ASKAP and ATCA are the only telescopes to have detected this SN\,2012dy at radio wavelengths because of its southern declination.

The emission revealed variability -- showing indications of multiple peaks -- with rapid evolution unlike any other radio SN observed so far. 
We also conducted further ATCA follow-up observations of SN\,2012dy as part of a targeted new monitoring campaign (C3442) to study this unique temporal and spectral evolution.  We split each of the three bands into four sub-bands ($512$\,MHz bandwidth) and repeated this imaging process in order to derive the spectral index  --- see Table \ref{tab:SED_SN2012dy}. 

\subsection{SN 2016coi}
\label{subsec: SN 2016coi}
SN\,2016coi was well studied in the year following its optical discovery \citep{2016ATel.9088....1G}. There are a number of ATCA observations during the first $\sim$200~days (ATOA Projects C3142, CX363) and several observations within the first $\sim274$\,days with the VLA. 

The VLA observations were conducted by \cite{2019ApJ...883..147T} whose subsequent analyses suggested a period of high mass-loss in the $\sim30$\,years prior to the explosion. They inferred a  shockwave moving at $v_{\rm sh}\sim0.15c$ which has since propagated into the dense surrounding medium.

\begin{figure}
\centering
\includegraphics[width=8.5cm]{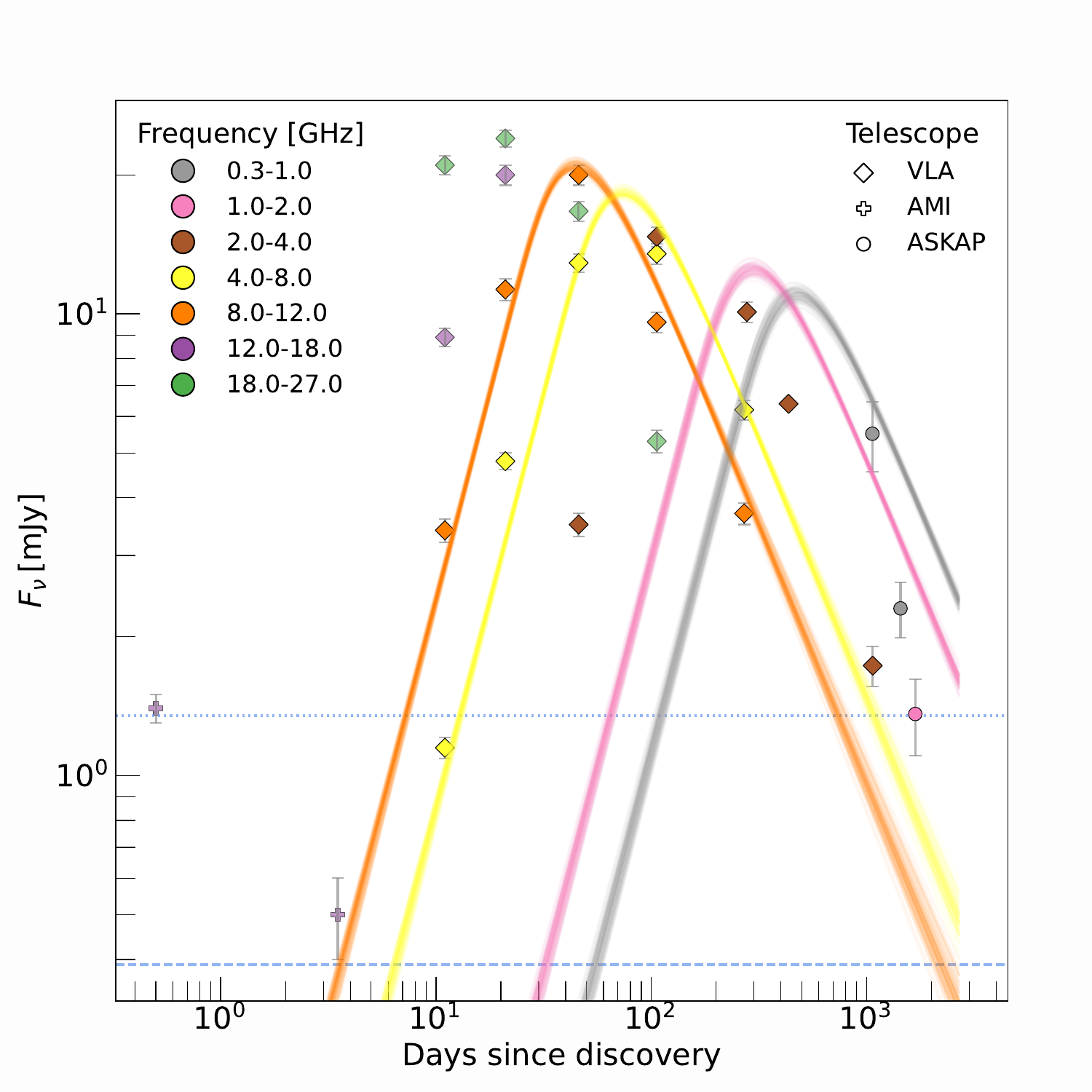}
\caption{Archival radio detections of SN\,2016coi \citep{2016ATel.9134....1M,2019ApJ...883..147T,2021ApJ...923...32B} along with the modelled lightcurves obtained from fitting of the early-time ($\Delta t<3\times10^3$\,days) data. We also plot detections from  all public ASKAP observations as of December 2023.  The MCMC parameterised lightcurves are extrapolated from the multi-frequency data and plotted for $0.8875, 1.4, 5.5, 9.0$\,GHz. The data binned into frequency bins and coloured accordingly.  We use diamond markers for VLA detections and circles for ASKAP. The dashed and dotted blue lines represents the FIRST and NVSS $3\sigma$ limits -- for any potential confusing host emission -- respectively.}
\label{fig: SN 2016coi lightcurve}
\end{figure}

SN\,2016coi was detected at late times by RACS-low and VLASS -- see Figure \ref{fig: SN 2016coi lightcurve} -- providing a complementary point of reference to see the decaying peak emission frequency. Like SN\,2003bg it is unclear whether there is a singular rising radio re-brightening event or if we are observing minor variability as part of secular lightcurve decay. While both are likely the result of the shockwave interaction with the CSM structure, significant CSM density enhancements or shells would be necessary to produce an order of magnitude rise in radio flux density.

\subsection{SN 2017gmr}
\label{subsec: SN 2017gmr}

SN\,2017gmr was discovered by \cite{2017ATel10706....1V} and classified as a SN II \citep{2017ATel10717....1P}. Photometric and  spectral analyses of early-time data found evidence of CSM interaction and an asymmetric SN explosion \citep{2019MNRAS.489L..69N,2019ApJ...885...43A}. There are no radio observations of SN\,2017gmr during the first $\sim$2\,years post-discovery.

\begin{figure}
\centering
\includegraphics[width=8.5cm]{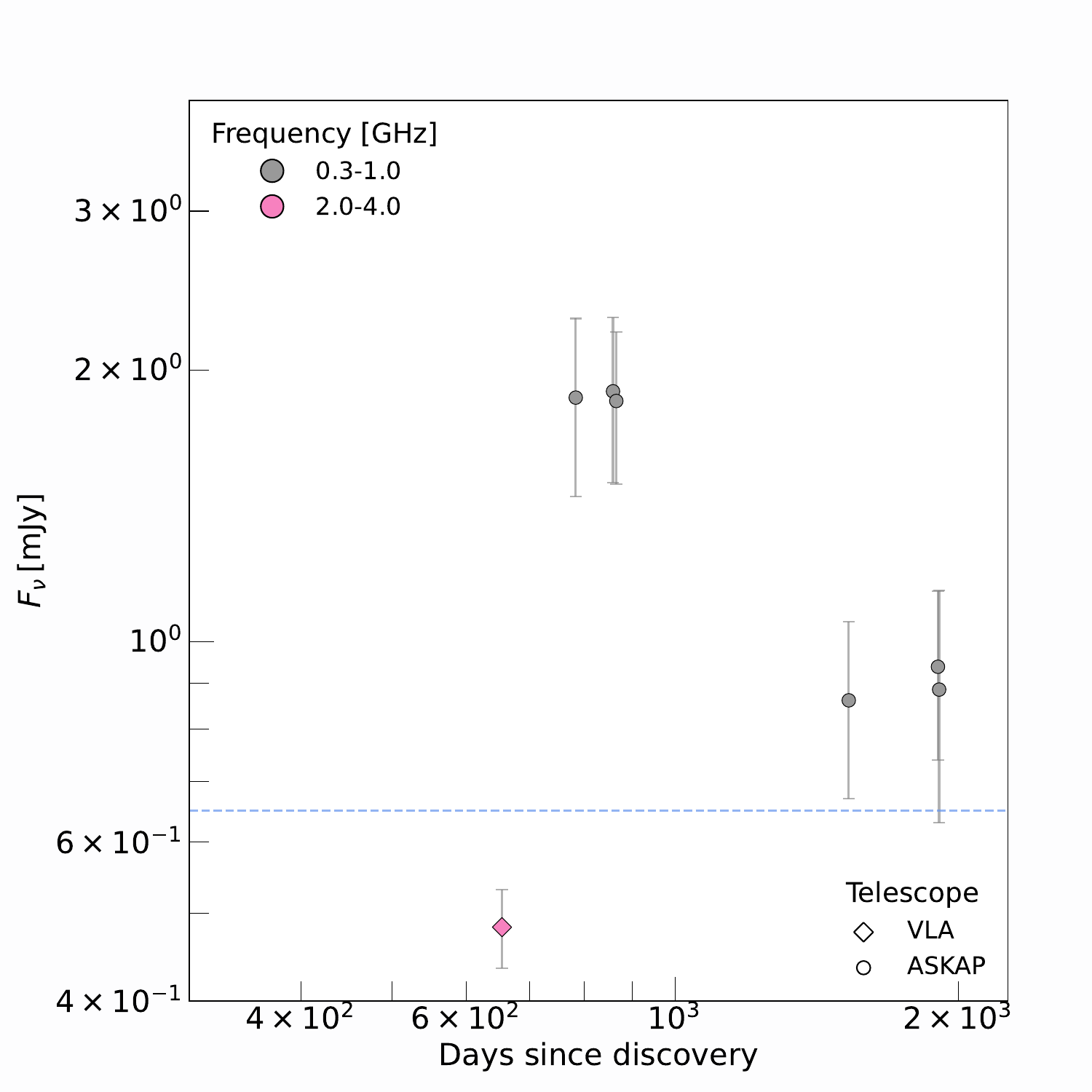}
\caption{Archival radio detections of SN\,2017gmr from the literature \citep{2020RNAAS...4..175G} and CASDA, including all public ASKAP observations as of December 2023, binned into frequency bins and coloured accordingly. We use diamond markers for VLA detections and circles for ASKAP. The dashed  blue line represents the FIRST $5\sigma$ limit for any potential confusing host emission.} 
\label{fig: SN 2017gmr Lightcurve Combined}
\end{figure}

The detection of radio emission at late-times could indicate a re-brightening event -- see Figure \ref{fig: SN 2017gmr Lightcurve Combined} -- but the emission appears to be consistent over nearly a year, as well as being quite close to the $5\sigma$ limit for both the FIRST and \texttt{Selavy} source detection thresholds.
Further observations are required to confirm whether there is any variability associated with this re-brightening. 
SN\,2017gmr provides a good example of the types of potential false positives which must be correctly identified by future variability surveys and transient detection pipelines.

\section{Discussion}
\label{sec: Discussion} 

\subsection{Variability \& Scintillation}
\label{sub-sec: Variability & Scintillation} 

The variability metrics shown in Equations \ref{eq:modulation index},\ref{eq: reduced chi-squared} are another way to identify false positives. We compared the calculated RISS variability indices, $m$, with the observed modulation metrics, $V$ and $\eta$, for these eight CCSNe --- summarised in Table \ref{scintillation table}. The variability metrics are calculated using the $887.5$\, MHz VAST data from $\Delta t > 10^3$\,days. We found that the observed late-time variability may not be the result of intrinsic variability of SN\,1996aq and SN\,2012ap, since $m\sim V$ and the signficance $\eta$ is low. For SN\,2003bg and SN\,2012dy the observed late-time variability may be the result of intrinsic source variability as $m<V$ and $1<\eta$. For these CCSNe the variability is likely due to the interaction of the shock with a complex CSM structure. Similarly, while the variability at $887.5$\, MHz may be explained by refractive scintillation, RISS cannot account for the order of magnitude re-brightening of SN\,2004dk at higher frequencies. This suggests that variability metrics alone cannot provide a definitive measure of whether or not a CCSNe is undergoing a period of late-time radio re-brightening. The variability of SN\,2016coi also appears to be the result of an intrinsic change in the SN, with the decline in flux density relative to the modelled emission suggesting a drop in CSM density. For SN\,2017gmr $m \gg V$ and the observed late-time emission is therefore consistent with extrinsic scintillation and not variability of the source, but more observations are required to confirm this.

\begin{table}
\centering 
\caption{The modulation index $V$ and the reduced $\chi^2$ metric $\eta$ calculated for the $887.5$ \,MHz ASKAP data. $m$ is the index of expected RISS variability at this frequency (see Section \ref{subsec: Modelling & Analysis}). We do not calculate the $V,\eta$ variability metrics for SN\,2013bi due to a paucity of detections. For SN\,2016coi there are only two $887.5$\,MHz ASKAP data points used to calculate these indices.}
\label{scintillation table}
\begin{tabular}{lccc}
\toprule
Name & $m$ & $V$ & $\eta$ \\
\midrule
SN 1996aq & 0.144 & 0.168 & 1.582 \\
SN 2003bg & 0.115 & 0.186 & 6.422 \\
SN 2004dk & 0.131 & 0.1 & 1.619 \\
SN 2012ap & 0.129 & 0.134 & 1.367 \\
SN 2012dy & 0.134 & 0.231 & 6.722 \\
SN 2013bi & 0.12 & -- & -- \\
SN 2016coi & 0.123 & 0.58 & 10.161 \\
SN 2017gmr & 0.139 & 0.012 & 0.004 \\
\bottomrule
\end{tabular}
\end{table}

\subsection{Implications \& Future Work}
\label{sub-sec: Implications & Future Work}

\begin{figure*}
\includegraphics[width=\textwidth]{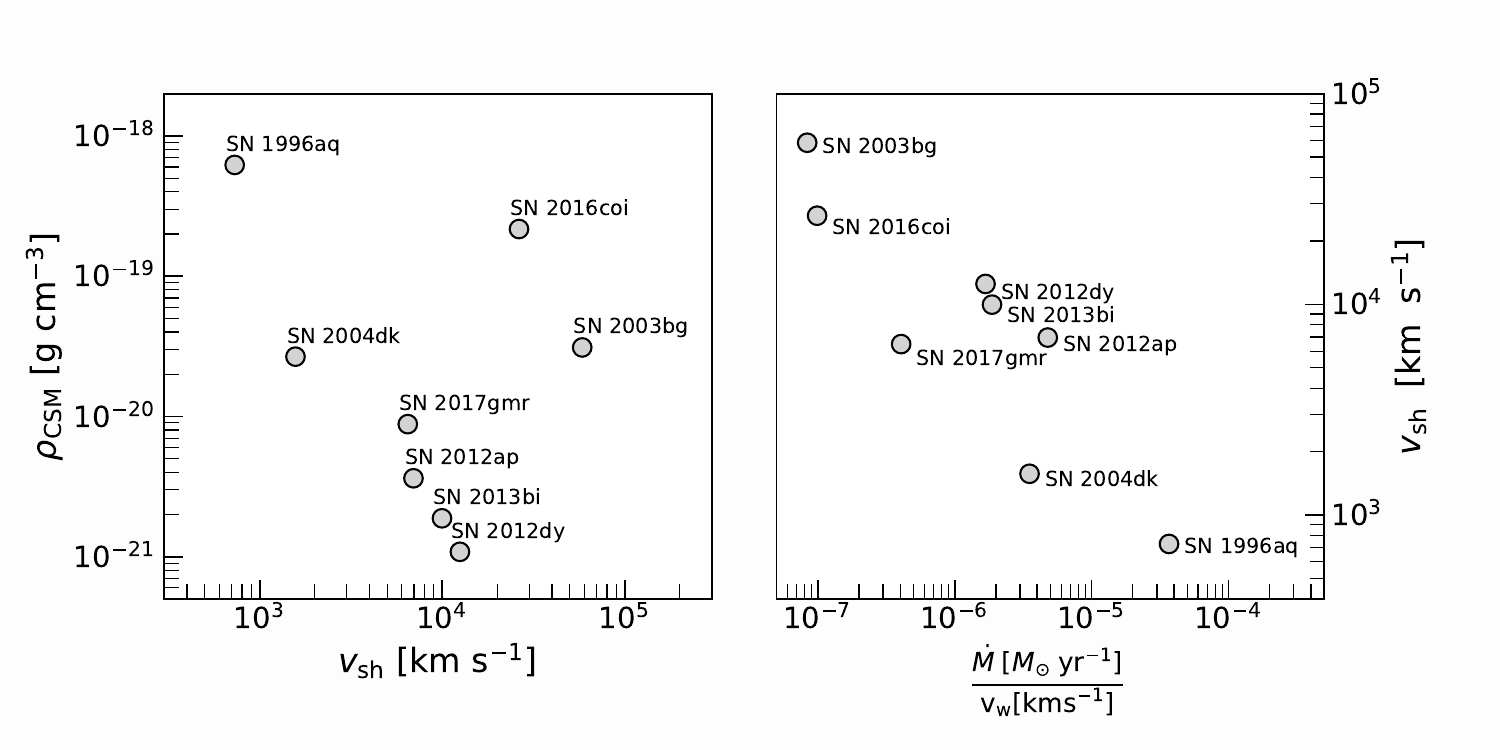}
  \caption{We calculate the CSM density and mass-loss rate with Equations \ref{eq: shock radius}, \ref{eq: magnetic field} for each CCSN at its respective peak flux density. \textit{Left}: CSM density $\rho_{\rm CSM}$ as a function of the shock velocity. \textit{Right}: Shock velocity compared to the mass-loss rate, noting that $\dot{M}$ is scaled by the wind velocities $10\,\rm{km~s^{-1}}$ and $1\,000\,\rm{km~s^{-1}}$ for Type II and Type Ib/c SNe, respectively}.
  \label{fig: Combined Shock Plots}
\end{figure*}

This work set out to find cases of late-time radio re-brightening in the pilot surveys VAST-P1 and RACS-low. These early results from the pilot phase have demonstrated the promising potential for ASKAP to find sources of late-time emission or place limits on late-time CSM interaction with ejected shells from earlier mass-loss episodes --- see Figure \ref{fig: Combined Shock Plots}. For example, we see that SN\,2003bg and SN\,2016coi have extremely fast outflow velocities, as expected for stripped-envelope SNe \citep{2021ApJ...908...75B}. But we also find that the mass-loss rates for SN\,2004dk and SN\,2012ap at late times are larger than predicted by \cite{2014ARA&A..52..487S}. This is a relatively unknown phase space that ASKAP is helping to fill. 

Detections of these CCSNe re-brightening events provide additional examples of dense CSM inhomogenities resulting from distinct periods of extreme mass-loss, breaking the standard model constant CSM density structure. While recent results continue to support the CSM-shock interaction interpretation of late-time SNe re-brightening \citep[e.g.][]{2022ApJ...938...84D,2024A&A...686A.129S}, alternative explanations -- like pulsar-wind nebulae or off-axis jets -- have also been suggested as central emission engines \citep[e.g.][]{2021ApJ...923L..24S,2023MNRAS.523.1474R}.

Future work on this topic will be aided by the VAST transient detection pipeline \citep{2022ASPC..532..333P}, which uses a streamlined source matching process across epochs to generate variability parameters used to identify sources of interest. Specifically, variables are identified based on the variability threshold and modulation index defined by \cite{2016ApJ...818..105M}.

Three SNe from our sample were included in this VAST pipeline and their variability metrics were reasonably consistent with our interpretation of the sources, namely: SN\,1996aq is undergoing a period of re-brightening which is slowly declining but difficult to model because the flux density variations are on the same order of magnitude as the errors. This source did not meet the pipeline's variability thresholds. SN\,2012dy is undergoing a period of re-brightening that is varying on larger scales with smaller errors, with the appearance of two distinct peaks followed by a gradual decline; and still appears to be undergoing some interaction. This source passes the pipeline's $95$ per cent variability confidence threshold but does not attain a sufficient modulation index; this only selects highly variable sources with an equivalent fractional flux density variation of three. While we do detect late-time radio emission from SN\,2017gmr, its variability appears to be consistent with scintillation around the Selavy $5\sigma$ cutoff. This source was not picked up by the \citeauthor{2022ASPC..532..333P} transient pipeline and is a good example of its ability to ignore a false positive.   

SN\,2013bi is another example of a false positive from our original selection process. This CCSN has almost no additional information besides its optical discovery \citep{2013CBET.3481....1J}. It was detected by RACS-low $\sim$6~years after discovery, and was faintly visible in VLASS several days later \citep{2020RNAAS...4..175G}, but SN\,2013bi has not been observed since. Leading up to the public release of the RACS-low data, \citet{2021PASA...38...58H} noted that a decreased resolution  -- due to an extended PSF in the observation of certain fields --  may have produced some erroneous \texttt{Selavy} detections of objects with neighbouring sources in our search. 

These fields were re-observed to create a reliable catalogue with more uniform resolution and SN\,2013bi was observed in one such field. In the updated observation, as in the more recent $1.365$\,GHz RACS survey \citep[RACS-mid;][]{2023PASA...40...34D}, there is significant emission from the host galaxy. Therefore we do not consider the original RACS-low flux density value reliable, as it may be confused with the extended emission. Given the sparsity of available radio detections we do not show a lightcurve or calculate the variability metrics for SN\,2013bi.

\section{Summary}
\label{sec: Summary} 
In this work we have demonstrated the general ability to detect late-time SN re-brightening events with VAST.

VAST-P1 observed the complete $5\,131$\,deg$^2$ pilot survey region four times and nominally identified $29$ CCSNe and five real late-time SNe re-brightening events. 
The full VAST survey  is observing a larger region of $\sim9\,500$\,deg$^2$ at an approximately bimonthly cadence over four years. A scaled estimate suggests that VAST will detect $\sim640$ CCSNe and $\sim110$ late-time SNe re-brightening events with luminosities of $L_{\nu}\sim10^{29}$\,erg s$^{-1}$Hz$^{-1}$ or greater. This estimate makes the simplifying assumption of uniform source density over the survey region.  

This approach may also be useful in identifying orphan afterglow events from long-duration GRBs, which are often observed to be accompanied by a Type Ic broad-lined CCSN (particularly in the local Universe). For a review of the GRB-SN connection, see \cite{2017AdAst2017E...5C}. 
While recent studies have shown that unbiased ASKAP surveys are now sensitive enough to detect radio emission from known GRBs \citep{2021MNRAS.503.1847L} and find strong orphan afterglow candidates \citep{2023MNRAS.523.4029L}, the discovery of late-time supernova-powered re-brightening expected in these GRB-SN events \citep[e.g.][]{2015MNRAS.454.1711B} remains elusive due to the scarcity of late-time (decades post-burst) follow-up. The method demonstrated in this work provides a promising avenue for discovering late-time re-brightenings from GRB-SN events, which has implications for our understanding of the progenitors of GRBs, orphan afterglow searches, and constraining their true population rate. 

VAST can also be used to identify and study late-time re-brightening in other classes of transients like TDEs \citep{2024arXiv240712097A}. Additionally, VAST can be used to monitor long-term variability in Active Galactic Nuclei, as well stellar emission \citep{pritchard_vast_stars_2024}.

\section*{Acknowledgements}

KR thanks the LSST-DA Data Science Fellowship Program, which is funded by LSST-DA, the Brinson Foundation, and the Moore Foundation; Their participation in the program has benefited this work. KR would like to acknowledge the help of Gemma E. Anderson and Lucinda Lilley for their helpful advice and insightful discussion. KR and RA also thank Joshua Pritchard for his help in the early stages of this project. AH is grateful for the support by the United States-Israel Binational Science Foundation (BSF grant 2020203) and by the Sir Zelman Cowen Universities Fund. This research was supported by the Israel Science Foundation (grant No. 1679/23). DLK was supported by NSF grant AST-1816492. 

We thank the Mullard Radio Astronomy Observatory (MRAO) for carrying out the Arcminute Micro-Kelvin Imager - Large Array (AMI-LA) observations.

The Australia Telescope Compact Array is part of the Australia Telescope National Facility which is funded by the Australian Government for operation as a National Facility managed by CSIRO. We acknowledge the Gomeroi people as the traditional owners of the Observatory site.

This scientific work uses data obtained from Inyarrimanha Ilgari Bundara / the Murchison Radio-astronomy Observatory. We acknowledge the Wajarri Yamaji People as the traditional owners of the Observatory site. CSIRO’s ASKAP radio telescope is part of the \href{https://ror.org/05qajvd42}{Australia Telescope National Facility}. Operation of ASKAP is funded by the Australian Government with support from the National Collaborative Research Infrastructure Strategy. ASKAP uses the resources of the Pawsey Supercomputing Research Centre. Establishment of ASKAP, Inyarrimanha Ilgari Bundara, the CSIRO Murchison Radio-astronomy Observatory and the Pawsey Supercomputing Research Centre are initiatives of the Australian Government, with support from the Government of Western Australia and the Science and Industry Endowment Fund.
This paper includes archived data obtained through the CSIRO ASKAP Science Data Archive, 
CASDA\footnote{\href{http://data.csiro.au/}{http://data.csiro.au/}}.

\section*{Data Availability}
A complete tabulation of all new and archival SN data used in this work is available in machine-readable form as an online supplement to this paper.

The ASKAP data used in this paper (RACS-low and VAST-P1) can be accessed through the CSIRO ASKAP Science Data Archive (CASDA) under project codes AS110 and AS107. The ATCA data used in this paper can be accessed through the Australia Telescope Online Archive (ATOA) under project codes C3363 and C3442. Other auxiliary datasets can be made available upon request via email to the corresponding author.



\bibliographystyle{mnras}
\bibliography{VAST_bibliography} 

\begin{thebibliography}{}
\makeatletter
\relax
\def\mn@urlcharsother{\let\do\@makeother \do\$\do\&\do\#\do\^\do\_\do\%\do\~}
\def\mn@doi{\begingroup\mn@urlcharsother \@ifnextchar [ {\mn@doi@} {\mn@doi@[]}}
\def\mn@doi@[#1]#2{\def\@tempa{#1}\ifx\@tempa\@empty \href {http://dx.doi.org/#2} {doi:#2}\else \href {http://dx.doi.org/#2} {#1}\fi \endgroup}
\def\mn@eprint#1#2{\mn@eprint@#1:#2::\@nil}
\def\mn@eprint@arXiv#1{\href {http://arxiv.org/abs/#1} {{\tt arXiv:#1}}}
\def\mn@eprint@dblp#1{\href {http://dblp.uni-trier.de/rec/bibtex/#1.xml} {dblp:#1}}
\def\mn@eprint@#1:#2:#3:#4\@nil{\def\@tempa {#1}\def\@tempb {#2}\def\@tempc {#3}\ifx \@tempc \@empty \let \@tempc \@tempb \let \@tempb \@tempa \fi \ifx \@tempb \@empty \def\@tempb {arXiv}\fi \@ifundefined {mn@eprint@\@tempb}{\@tempb:\@tempc}{\expandafter \expandafter \csname mn@eprint@\@tempb\endcsname \expandafter{\@tempc}}}

\bibitem[\protect\citeauthoryear{{Anderson} et~al.,}{{Anderson} et~al.}{2017}]{2017MNRAS.466.3648A}
{Anderson} G.~E.,  et~al., 2017, \mn@doi [\mnras] {10.1093/mnras/stw3310}, \href {https://ui.adsabs.harvard.edu/abs/2017MNRAS.466.3648A} {466, 3648}

\bibitem[\protect\citeauthoryear{{Andrews} et~al.,}{{Andrews} et~al.}{2019}]{2019ApJ...885...43A}
{Andrews} J.~E.,  et~al., 2019, \mn@doi [\apj] {10.3847/1538-4357/ab43e3}, \href {https://ui.adsabs.harvard.edu/abs/2019ApJ...885...43A} {885, 43}

\bibitem[\protect\citeauthoryear{{Anumarlapudi} et~al.,}{{Anumarlapudi} et~al.}{2024}]{2024arXiv240712097A}
{Anumarlapudi} A.,  et~al., 2024, arXiv e-prints, \href {https://ui.adsabs.harvard.edu/abs/2024arXiv240712097A} {p. arXiv:2407.12097}

\bibitem[\protect\citeauthoryear{{Balasubramanian}, {Corsi}, {Polisensky}, {Clarke}  \& {Kassim}}{{Balasubramanian} et~al.}{2021}]{2021ApJ...923...32B}
{Balasubramanian} A.,  {Corsi} A.,  {Polisensky} E.,  {Clarke} T.~E.,   {Kassim} N.~E.,  2021, \mn@doi [\apj] {10.3847/1538-4357/ac2154}, \href {https://ui.adsabs.harvard.edu/abs/2021ApJ...923...32B} {923, 32}

\bibitem[\protect\citeauthoryear{{Barbon}, {Buond{\'\i}}, {Cappellaro}  \& {Turatto}}{{Barbon} et~al.}{1999}]{1999A&AS..139..531B}
{Barbon} R.,  {Buond{\'\i}} V.,  {Cappellaro} E.,   {Turatto} M.,  1999, \mn@doi [\aaps] {10.1051/aas:1999404}, \href {https://ui.adsabs.harvard.edu/abs/1999A&AS..139..531B} {139, 531}

\bibitem[\protect\citeauthoryear{{Barniol Duran} \& {Giannios}}{{Barniol Duran} \& {Giannios}}{2015}]{2015MNRAS.454.1711B}
{Barniol Duran} R.,  {Giannios} D.,  2015, \mn@doi [\mnras] {10.1093/mnras/stv2004}, \href {https://ui.adsabs.harvard.edu/abs/2015MNRAS.454.1711B} {454, 1711}

\bibitem[\protect\citeauthoryear{{Becker}, {White}  \& {Helfand}}{{Becker} et~al.}{1995}]{1995ApJ...450..559B}
{Becker} R.~H.,  {White} R.~L.,   {Helfand} D.~J.,  1995, \mn@doi [\apj] {10.1086/176166}, \href {https://ui.adsabs.harvard.edu/abs/1995ApJ...450..559B} {450, 559}

\bibitem[\protect\citeauthoryear{{Bietenholz}, {Bartel}, {Argo}, {Dua}, {Ryder}  \& {Soderberg}}{{Bietenholz} et~al.}{2021}]{2021ApJ...908...75B}
{Bietenholz} M.~F.,  {Bartel} N.,  {Argo} M.,  {Dua} R.,  {Ryder} S.,   {Soderberg} A.,  2021, \mn@doi [\apj] {10.3847/1538-4357/abccd9}, \href {https://ui.adsabs.harvard.edu/abs/2021ApJ...908...75B} {908, 75}

\bibitem[\protect\citeauthoryear{{Bock}, {Milisavljevic}, {Parrent}, {Fesen}, {Margutti}, {Soderberg}, {Pickering}  \& {Kotze}}{{Bock} et~al.}{2012}]{2012CBET.3197....1B}
{Bock} G.,  {Milisavljevic} D.,  {Parrent} J.~T.,  {Fesen} R.,  {Margutti} R.,  {Soderberg} A.,  {Pickering} T.,   {Kotze} P.,  2012, Central Bureau Electronic Telegrams, \href {https://ui.adsabs.harvard.edu/abs/2012CBET.3197....1B} {3197}

\bibitem[\protect\citeauthoryear{{Cano}, {Wang}, {Dai}  \& {Wu}}{{Cano} et~al.}{2017}]{2017AdAst2017E...5C}
{Cano} Z.,  {Wang} S.-Q.,  {Dai} Z.-G.,   {Wu} X.-F.,  2017, \mn@doi [Advances in Astronomy] {10.1155/2017/8929054}, \href {https://ui.adsabs.harvard.edu/abs/2017AdAst2017E...5C} {2017, 8929054}

\bibitem[\protect\citeauthoryear{{Chakraborti}, {Chandra}  \& {Ray}}{{Chakraborti} et~al.}{2009}]{2009ATel.1974....1C}
{Chakraborti} S.,  {Chandra} P.,   {Ray} A.,  2009, The Astronomer's Telegram, \href {https://ui.adsabs.harvard.edu/abs/2009ATel.1974....1C} {1974}

\bibitem[\protect\citeauthoryear{{Chakraborti} et~al.,}{{Chakraborti} et~al.}{2015}]{2015ApJ...805..187C}
{Chakraborti} S.,  et~al., 2015, \mn@doi [\apj] {10.1088/0004-637X/805/2/187}, \href {https://ui.adsabs.harvard.edu/abs/2015ApJ...805..187C} {805, 187}

\bibitem[\protect\citeauthoryear{{Chevalier}}{{Chevalier}}{1982}]{1982ApJ...259..302C}
{Chevalier} R.~A.,  1982, \mn@doi [\apj] {10.1086/160167}, \href {https://ui.adsabs.harvard.edu/abs/1982ApJ...259..302C} {259, 302}

\bibitem[\protect\citeauthoryear{{Chevalier}}{{Chevalier}}{1996}]{1996ASPC...93..125C}
{Chevalier} R.~A.,  1996, in {Taylor} A.~R.,  {Paredes} J.~M.,  eds,  Astronomical Society of the Pacific Conference Series Vol. 93, Radio Emission from the Stars and the Sun. p.~125

\bibitem[\protect\citeauthoryear{{Chevalier}}{{Chevalier}}{1998}]{1998ApJ...499..810C}
{Chevalier} R.~A.,  1998, \mn@doi [\apj] {10.1086/305676}, \href {https://ui.adsabs.harvard.edu/abs/1998ApJ...499..810C} {499, 810}

\bibitem[\protect\citeauthoryear{{Clarke}, {Kassim}, {Brisken}, {Helmboldt}, {Peters}, {Ray}, {Polisensky}  \& {Giacintucci}}{{Clarke} et~al.}{2016}]{2016SPIE.9906E..5BC}
{Clarke} T.~E.,  {Kassim} N.~E.,  {Brisken} W.,  {Helmboldt} J.,  {Peters} W.,  {Ray} P.~S.,  {Polisensky} E.,   {Giacintucci} S.,  2016, in {Hall} H.~J.,  {Gilmozzi} R.,   {Marshall} H.~K.,  eds,  Society of Photo-Optical Instrumentation Engineers (SPIE) Conference Series Vol. 9906, Ground-based and Airborne Telescopes VI. p. 99065B, \mn@doi{10.1117/12.2233036}

\bibitem[\protect\citeauthoryear{{Condon}, {Cotton}, {Greisen}, {Yin}, {Perley}, {Taylor}  \& {Broderick}}{{Condon} et~al.}{1998}]{1998AJ....115.1693C}
{Condon} J.~J.,  {Cotton} W.~D.,  {Greisen} E.~W.,  {Yin} Q.~F.,  {Perley} R.~A.,  {Taylor} G.~B.,   {Broderick} J.~J.,  1998, \mn@doi [\aj] {10.1086/300337}, \href {https://ui.adsabs.harvard.edu/abs/1998AJ....115.1693C} {115, 1693}

\bibitem[\protect\citeauthoryear{{Cornwell}, {Voronkov}  \& {Humphreys}}{{Cornwell} et~al.}{2012}]{2012SPIE.8500E..0LC}
{Cornwell} T.~J.,  {Voronkov} M.~A.,   {Humphreys} B.,  2012, in {Bones} P.~J.,  {Fiddy} M.~A.,   {Millane} R.~P.,  eds,  Society of Photo-Optical Instrumentation Engineers (SPIE) Conference Series Vol. 8500, Image Reconstruction from Incomplete Data VII. p. 85000L (\mn@eprint {arXiv} {1207.5861}), \mn@doi{10.1117/12.929336}

\bibitem[\protect\citeauthoryear{{DeMarchi} et~al.,}{{DeMarchi} et~al.}{2022}]{2022ApJ...938...84D}
{DeMarchi} L.,  et~al., 2022, \mn@doi [\apj] {10.3847/1538-4357/ac8c26}, \href {https://ui.adsabs.harvard.edu/abs/2022ApJ...938...84D} {938, 84}

\bibitem[\protect\citeauthoryear{{Duchesne} et~al.,}{{Duchesne} et~al.}{2023}]{2023PASA...40...34D}
{Duchesne} S.~W.,  et~al., 2023, \mn@doi [\pasa] {10.1017/pasa.2023.31}, \href {https://ui.adsabs.harvard.edu/abs/2023PASA...40...34D} {40, e034}

\bibitem[\protect\citeauthoryear{{Foreman-Mackey}, {Hogg}, {Lang}  \& {Goodman}}{{Foreman-Mackey} et~al.}{2013}]{2013PASP..125..306F}
{Foreman-Mackey} D.,  {Hogg} D.~W.,  {Lang} D.,   {Goodman} J.,  2013, \mn@doi [\pasp] {10.1086/670067}, \href {https://ui.adsabs.harvard.edu/abs/2013PASP..125..306F} {125, 306}

\bibitem[\protect\citeauthoryear{{Frieman} et~al.,}{{Frieman} et~al.}{2008}]{2008AJ....135..338F}
{Frieman} J.~A.,  et~al., 2008, \mn@doi [\aj] {10.1088/0004-6256/135/1/338}, \href {https://ui.adsabs.harvard.edu/abs/2008AJ....135..338F} {135, 338}

\bibitem[\protect\citeauthoryear{{Gal-Yam}}{{Gal-Yam}}{2021}]{2021AAS...23742305G}
{Gal-Yam} A.,  2021, in American Astronomical Society Meeting Abstracts. p. 423.05

\bibitem[\protect\citeauthoryear{{Gordon} et~al.,}{{Gordon} et~al.}{2020}]{2020RNAAS...4..175G}
{Gordon} Y.~A.,  et~al., 2020, \mn@doi [Research Notes of the American Astronomical Society] {10.3847/2515-5172/abbe23}, \href {https://ui.adsabs.harvard.edu/abs/2020RNAAS...4..175G} {4, 175}

\bibitem[\protect\citeauthoryear{{Grupe}, {Brown}, {Dong}, {Shappee}, {Holoien}, {Stanek}, {Prieto}  \& {Margutti}}{{Grupe} et~al.}{2016}]{2016ATel.9088....1G}
{Grupe} D.,  {Brown} P.,  {Dong} S.,  {Shappee} B.~J.,  {Holoien} T.,  {Stanek} K.,  {Prieto} J.~L.,   {Margutti} R.,  2016, The Astronomer's Telegram, \href {https://ui.adsabs.harvard.edu/abs/2016ATel.9088....1G} {9088}

\bibitem[\protect\citeauthoryear{{Guillochon}, {Parrent}, {Kelley}  \& {Margutti}}{{Guillochon} et~al.}{2017}]{2017ApJ...835...64G}
{Guillochon} J.,  {Parrent} J.,  {Kelley} L.~Z.,   {Margutti} R.,  2017, \mn@doi [\apj] {10.3847/1538-4357/835/1/64}, \href {https://ui.adsabs.harvard.edu/abs/2017ApJ...835...64G} {835, 64}

\bibitem[\protect\citeauthoryear{{Hale} et~al.,}{{Hale} et~al.}{2021}]{2021PASA...38...58H}
{Hale} C.~L.,  et~al., 2021, \mn@doi [\pasa] {10.1017/pasa.2021.47}, \href {https://ui.adsabs.harvard.edu/abs/2021PASA...38...58H} {38, e058}

\bibitem[\protect\citeauthoryear{{Hamuy} et~al.,}{{Hamuy} et~al.}{2009}]{2009ApJ...703.1612H}
{Hamuy} M.,  et~al., 2009, \mn@doi [\apj] {10.1088/0004-637X/703/2/1612}, \href {https://ui.adsabs.harvard.edu/abs/2009ApJ...703.1612H} {703, 1612}

\bibitem[\protect\citeauthoryear{{Hancock}, {Charlton}, {Macquart}  \& {Hurley-Walker}}{{Hancock} et~al.}{2019}]{2019arXiv190708395H}
{Hancock} P.~J.,  {Charlton} E.~G.,  {Macquart} J.-P.,   {Hurley-Walker} N.,  2019, arXiv e-prints, \href {https://ui.adsabs.harvard.edu/abs/2019arXiv190708395H} {p. arXiv:1907.08395}

\bibitem[\protect\citeauthoryear{{Heger}, {Fryer}, {Woosley}, {Langer}  \& {Hartmann}}{{Heger} et~al.}{2003}]{2003ApJ...591..288H}
{Heger} A.,  {Fryer} C.~L.,  {Woosley} S.~E.,  {Langer} N.,   {Hartmann} D.~H.,  2003, \mn@doi [\apj] {10.1086/375341}, \href {https://ui.adsabs.harvard.edu/abs/2003ApJ...591..288H} {591, 288}

\bibitem[\protect\citeauthoryear{Hickish et~al.,}{Hickish et~al.}{2018}]{hickish_2018}
Hickish J.,  et~al., 2018, \mn@doi [\mnras] {10.1093/mnras/sty074}, 475, 5677

\bibitem[\protect\citeauthoryear{{Horesh} et~al.,}{{Horesh} et~al.}{2020}]{2020ApJ...903..132H}
{Horesh} A.,  et~al., 2020, \mn@doi [\apj] {10.3847/1538-4357/abbd38}, \href {https://ui.adsabs.harvard.edu/abs/2020ApJ...903..132H} {903, 132}

\bibitem[\protect\citeauthoryear{{Hotan} et~al.,}{{Hotan} et~al.}{2021}]{2021PASA...38....9H}
{Hotan} A.~W.,  et~al., 2021, \mn@doi [\pasa] {10.1017/pasa.2021.1}, \href {https://ui.adsabs.harvard.edu/abs/2021PASA...38....9H} {38, e009}

\bibitem[\protect\citeauthoryear{{Jewett} et~al.,}{{Jewett} et~al.}{2012}]{2012CBET.3037....1J}
{Jewett} L.,  et~al., 2012, Central Bureau Electronic Telegrams, \href {https://ui.adsabs.harvard.edu/abs/2012CBET.3037....1J} {3037}

\bibitem[\protect\citeauthoryear{{Jin} et~al.,}{{Jin} et~al.}{2013}]{2013CBET.3481....1J}
{Jin} Z.,  et~al., 2013, Central Bureau Electronic Telegrams, \href {https://ui.adsabs.harvard.edu/abs/2013CBET.3481....1J} {3481}

\bibitem[\protect\citeauthoryear{Lacy et~al.,}{Lacy et~al.}{2020}]{Lacy_2020}
Lacy M.,  et~al., 2020, \mn@doi [PASP] {10.1088/1538-3873/ab63eb}, 132, 035001

\bibitem[\protect\citeauthoryear{{Langer}}{{Langer}}{2012}]{2012ARA&A..50..107L}
{Langer} N.,  2012, \mn@doi [\araa] {10.1146/annurev-astro-081811-125534}, \href {https://ui.adsabs.harvard.edu/abs/2012ARA&A..50..107L} {50, 107}

\bibitem[\protect\citeauthoryear{{Leung} et~al.,}{{Leung} et~al.}{2021}]{2021MNRAS.503.1847L}
{Leung} J.~K.,  et~al., 2021, \mn@doi [\mnras] {10.1093/mnras/stab326}, \href {https://ui.adsabs.harvard.edu/abs/2021MNRAS.503.1847L} {503, 1847}

\bibitem[\protect\citeauthoryear{{Leung}, {Murphy}, {Lenc}, {Edwards}, {Ghirlanda}, {Kaplan}, {O'Brien}  \& {Wang}}{{Leung} et~al.}{2023}]{2023MNRAS.523.4029L}
{Leung} J.~K.,  {Murphy} T.,  {Lenc} E.,  {Edwards} P.~G.,  {Ghirlanda} G.,  {Kaplan} D.~L.,  {O'Brien} A.,   {Wang} Z.,  2023, \mn@doi [\mnras] {10.1093/mnras/stad1670}, \href {https://ui.adsabs.harvard.edu/abs/2023MNRAS.523.4029L} {523, 4029}

\bibitem[\protect\citeauthoryear{{Margalit} \& {Quataert}}{{Margalit} \& {Quataert}}{2024}]{2024arXiv240307048M}
{Margalit} B.,  {Quataert} E.,  2024, \mn@doi [arXiv e-prints] {10.48550/arXiv.2403.07048}, \href {https://ui.adsabs.harvard.edu/abs/2024arXiv240307048M} {p. arXiv:2403.07048}

\bibitem[\protect\citeauthoryear{{Mauerhan}, {Filippenko}, {Zheng}, {Brink}, {Graham}, {Shivvers}  \& {Clubb}}{{Mauerhan} et~al.}{2018}]{2018MNRAS.478.5050M}
{Mauerhan} J.~C.,  {Filippenko} A.~V.,  {Zheng} W.,  {Brink} T.~G.,  {Graham} M.~L.,  {Shivvers} I.,   {Clubb} K.~I.,  2018, \mn@doi [\mnras] {10.1093/mnras/sty1307}, \href {https://ui.adsabs.harvard.edu/abs/2018MNRAS.478.5050M} {478, 5050}

\bibitem[\protect\citeauthoryear{{Mazzali}, {Deng}, {Hamuy}  \& {Nomoto}}{{Mazzali} et~al.}{2009}]{2009ApJ...703.1624M}
{Mazzali} P.~A.,  {Deng} J.,  {Hamuy} M.,   {Nomoto} K.,  2009, \mn@doi [\apj] {10.1088/0004-637X/703/2/1624}, \href {https://ui.adsabs.harvard.edu/abs/2009ApJ...703.1624M} {703, 1624}

\bibitem[\protect\citeauthoryear{{McConnell} et~al.,}{{McConnell} et~al.}{2020}]{2020PASA...37...48M}
{McConnell} D.,  et~al., 2020, \mn@doi [\pasa] {10.1017/pasa.2020.41}, \href {https://ui.adsabs.harvard.edu/abs/2020PASA...37...48M} {37, e048}

\bibitem[\protect\citeauthoryear{{McMullin}, {Waters}, {Schiebel}, {Young}  \& {Golap}}{{McMullin} et~al.}{2007}]{2007ASPC..376..127M}
{McMullin} J.~P.,  {Waters} B.,  {Schiebel} D.,  {Young} W.,   {Golap} K.,  2007, in {Shaw} R.~A.,  {Hill} F.,   {Bell} D.~J.,  eds,  Astronomical Society of the Pacific Conference Series Vol. 376, Astronomical Data Analysis Software and Systems XVI. p.~127

\bibitem[\protect\citeauthoryear{{Milisavljevic} et~al.,}{{Milisavljevic} et~al.}{2013}]{2013ApJ...767...71M}
{Milisavljevic} D.,  et~al., 2013, \mn@doi [\apj] {10.1088/0004-637X/767/1/71}, \href {https://ui.adsabs.harvard.edu/abs/2013ApJ...767...71M} {767, 71}

\bibitem[\protect\citeauthoryear{{Milisavljevic} et~al.,}{{Milisavljevic} et~al.}{2015}]{2015ApJ...799...51M}
{Milisavljevic} D.,  et~al., 2015, \mn@doi [\apj] {10.1088/0004-637X/799/1/51}, \href {https://ui.adsabs.harvard.edu/abs/2015ApJ...799...51M} {799, 51}

\bibitem[\protect\citeauthoryear{{Mooley} et~al.,}{{Mooley} et~al.}{2016a}]{2016ApJ...818..105M}
{Mooley} K.~P.,  et~al., 2016a, \mn@doi [\apj] {10.3847/0004-637X/818/2/105}, \href {https://ui.adsabs.harvard.edu/abs/2016ApJ...818..105M} {818, 105}

\bibitem[\protect\citeauthoryear{{Mooley} et~al.,}{{Mooley} et~al.}{2016b}]{2016ATel.9134....1M}
{Mooley} K.~P.,  et~al., 2016b, The Astronomer's Telegram, \href {https://ui.adsabs.harvard.edu/abs/2016ATel.9134....1M} {9134}

\bibitem[\protect\citeauthoryear{{Moriya}, {Groh}  \& {Meynet}}{{Moriya} et~al.}{2013}]{2013A&A...557L...2M}
{Moriya} T.~J.,  {Groh} J.~H.,   {Meynet} G.,  2013, \mn@doi [\aap] {10.1051/0004-6361/201322012}, \href {https://ui.adsabs.harvard.edu/abs/2013A&A...557L...2M} {557, L2}

\bibitem[\protect\citeauthoryear{{Murphy} et~al.,}{{Murphy} et~al.}{2013}]{2013PASA...30....6M}
{Murphy} T.,  et~al., 2013, \mn@doi [\pasa] {10.1017/pasa.2012.006}, \href {https://ui.adsabs.harvard.edu/abs/2013PASA...30....6M} {30, e006}

\bibitem[\protect\citeauthoryear{{Murphy} et~al.,}{{Murphy} et~al.}{2021}]{2021PASA...38...54M}
{Murphy} T.,  et~al., 2021, \mn@doi [\pasa] {10.1017/pasa.2021.44}, \href {https://ui.adsabs.harvard.edu/abs/2021PASA...38...54M} {38, e054}

\bibitem[\protect\citeauthoryear{{Nagao} et~al.,}{{Nagao} et~al.}{2019}]{2019MNRAS.489L..69N}
{Nagao} T.,  et~al., 2019, \mn@doi [\mnras] {10.1093/mnrasl/slz119}, \href {https://ui.adsabs.harvard.edu/abs/2019MNRAS.489L..69N} {489, L69}

\bibitem[\protect\citeauthoryear{{Nakano} et~al.,}{{Nakano} et~al.}{1996}]{1996IAUC.6454....1N}
{Nakano} S.,  et~al., 1996, \iaucirc, \href {https://ui.adsabs.harvard.edu/abs/1996IAUC.6454....1N} {6454}

\bibitem[\protect\citeauthoryear{{Pacholczyk}}{{Pacholczyk}}{1970}]{1970ranp.book.....P}
{Pacholczyk} A.~G.,  1970, {Radio astrophysics. Nonthermal processes in galactic and extragalactic sources}.
W. H. Freeman

\bibitem[\protect\citeauthoryear{{Palliyaguru}, {Corsi}, {P{\'e}rez-Torres}, {Varenius}  \& {Van Eerten}}{{Palliyaguru} et~al.}{2021}]{2021ApJ...910...16P}
{Palliyaguru} N.~T.,  {Corsi} A.,  {P{\'e}rez-Torres} M.,  {Varenius} E.,   {Van Eerten} H.,  2021, \mn@doi [\apj] {10.3847/1538-4357/abe1c9}, \href {https://ui.adsabs.harvard.edu/abs/2021ApJ...910...16P} {910, 16}

\bibitem[\protect\citeauthoryear{{Perez-Torres} et~al.,}{{Perez-Torres} et~al.}{2015}]{2015Perez-Torres}
{Perez-Torres} M.,  et~al., 2015, in Advancing Astrophysics with the Square Kilometre Array (AASKA14). p.~60 (\mn@eprint {arXiv} {1409.1827})

\bibitem[\protect\citeauthoryear{{Perley}, {Chandler}, {Butler}  \& {Wrobel}}{{Perley} et~al.}{2011}]{perley2011}
{Perley} R.~A.,  {Chandler} C.~J.,  {Butler} B.~J.,   {Wrobel} J.~M.,  2011, \mn@doi [\apjl] {10.1088/2041-8205/739/1/L1}, \href {https://ui.adsabs.harvard.edu/abs/2011ApJ...739L...1P} {739, L1}

\bibitem[\protect\citeauthoryear{Perrott et~al.,}{Perrott et~al.}{2013}]{perrott_2013}
Perrott Y.~C.,  et~al., 2013, \mn@doi [\mnras] {10.1093/mnras/sts589}, 429, 3330

\bibitem[\protect\citeauthoryear{{Pintaldi}, {Stewart}, {O'Brien}, {Kaplan}  \& {Murphy}}{{Pintaldi} et~al.}{2022}]{2022ASPC..532..333P}
{Pintaldi} S.,  {Stewart} A.,  {O'Brien} A.,  {Kaplan} D.,   {Murphy} T.,  2022, in {Ruiz} J.~E.,  {Pierfedereci} F.,   {Teuben} P.,  eds,  Astronomical Society of the Pacific Conference Series Vol. 532, Astronomical Society of the Pacific Conference Series. p.~333 (\mn@eprint {arXiv} {2101.05898}), \mn@doi{10.48550/arXiv.2101.05898}

\bibitem[\protect\citeauthoryear{{Prentice} et~al.,}{{Prentice} et~al.}{2016}]{2016MNRAS.458.2973P}
{Prentice} S.~J.,  et~al., 2016, \mn@doi [\mnras] {10.1093/mnras/stw299}, \href {https://ui.adsabs.harvard.edu/abs/2016MNRAS.458.2973P} {458, 2973}

\bibitem[\protect\citeauthoryear{{Pritchard}, {Murphy}, {Heald}, {Wheatland}, {Kaplan}, {Lenc}, {O'Brien}  \& {Wang}}{{Pritchard} et~al.}{2024}]{pritchard_vast_stars_2024}
{Pritchard} J.,  {Murphy} T.,  {Heald} G.,  {Wheatland} M.~S.,  {Kaplan} D.~L.,  {Lenc} E.,  {O'Brien} A.,   {Wang} Z.,  2024, \mn@doi [\mnras] {10.1093/mnras/stae127}, \href {https://ui.adsabs.harvard.edu/abs/2024MNRAS.529.1258P} {529, 1258}

\bibitem[\protect\citeauthoryear{{Pursimo} et~al.,}{{Pursimo} et~al.}{2017}]{2017ATel10717....1P}
{Pursimo} T.,  et~al., 2017, The Astronomer's Telegram, \href {https://ui.adsabs.harvard.edu/abs/2017ATel10717....1P} {10717, 1}

\bibitem[\protect\citeauthoryear{{Rajala} et~al.,}{{Rajala} et~al.}{2005}]{2005PASP..117..132R}
{Rajala} A.~M.,  et~al., 2005, \mn@doi [\pasp] {10.1086/427985}, \href {https://ui.adsabs.harvard.edu/abs/2005PASP..117..132R} {117, 132}

\bibitem[\protect\citeauthoryear{{Rizzo Smith}, {Kochanek}  \& {Neustadt}}{{Rizzo Smith} et~al.}{2023}]{2023MNRAS.523.1474R}
{Rizzo Smith} M.,  {Kochanek} C.~S.,   {Neustadt} J.~M.~M.,  2023, \mn@doi [\mnras] {10.1093/mnras/stad1483}, \href {https://ui.adsabs.harvard.edu/abs/2023MNRAS.523.1474R} {523, 1474}

\bibitem[\protect\citeauthoryear{{Ruiz-Carmona}, {Sfaradi}  \& {Horesh}}{{Ruiz-Carmona} et~al.}{2022}]{2022A&A...666A..82R}
{Ruiz-Carmona} R.,  {Sfaradi} I.,   {Horesh} A.,  2022, \mn@doi [\aap] {10.1051/0004-6361/202142024}, \href {https://ui.adsabs.harvard.edu/abs/2022A&A...666A..82R} {666, A82}

\bibitem[\protect\citeauthoryear{{Ryder}, {Sadler}, {Subrahmanyan}, {Weiler}, {Panagia}  \& {Stockdale}}{{Ryder} et~al.}{2004}]{Ryder2004}
{Ryder} S.~D.,  {Sadler} E.~M.,  {Subrahmanyan} R.,  {Weiler} K.~W.,  {Panagia} N.,   {Stockdale} C.,  2004, \mn@doi [\mnras] {10.1111/j.1365-2966.2004.07589.x}, \href {https://ui.adsabs.harvard.edu/abs/2004MNRAS.349.1093R} {349, 1093}

\bibitem[\protect\citeauthoryear{{Salas}, {Bauer}, {Stockdale}  \& {Prieto}}{{Salas} et~al.}{2013}]{2013MNRAS.428.1207S}
{Salas} P.,  {Bauer} F.~E.,  {Stockdale} C.,   {Prieto} J.~L.,  2013, \mn@doi [\mnras] {10.1093/mnras/sts104}, \href {https://ui.adsabs.harvard.edu/abs/2013MNRAS.428.1207S} {428, 1207}

\bibitem[\protect\citeauthoryear{{Sault}, {Teuben}  \& {Wright}}{{Sault} et~al.}{1995}]{1995ASPC...77..433S}
{Sault} R.~J.,  {Teuben} P.~J.,   {Wright} M.~C.~H.,  1995, in {Shaw} R.~A.,  {Payne} H.~E.,   {Hayes} J.~J.~E.,  eds,  Astronomical Society of the Pacific Conference Series Vol. 77, Astronomical Data Analysis Software and Systems IV. p.~433 (\mn@eprint {arXiv} {astro-ph/0612759})

\bibitem[\protect\citeauthoryear{{Sfaradi}, {Horesh}, {Fender}, {Green}, {Williams}, {Bright}  \& {Schulze}}{{Sfaradi} et~al.}{2022}]{2022ApJ...933..176S}
{Sfaradi} I.,  {Horesh} A.,  {Fender} R.,  {Green} D.~A.,  {Williams} D. R.~A.,  {Bright} J.,   {Schulze} S.,  2022, \mn@doi [\apj] {10.3847/1538-4357/ac74bc}, \href {https://ui.adsabs.harvard.edu/abs/2022ApJ...933..176S} {933, 176}

\bibitem[\protect\citeauthoryear{{Sfaradi} et~al.,}{{Sfaradi} et~al.}{2024}]{2024A&A...686A.129S}
{Sfaradi} I.,  et~al., 2024, \mn@doi [\aap] {10.1051/0004-6361/202348761}, \href {https://ui.adsabs.harvard.edu/abs/2024A&A...686A.129S} {686, A129}

\bibitem[\protect\citeauthoryear{{Smartt} et~al.,}{{Smartt} et~al.}{2015}]{2015A&A...579A..40S}
{Smartt} S.~J.,  et~al., 2015, \mn@doi [\aap] {10.1051/0004-6361/201425237}, \href {https://ui.adsabs.harvard.edu/abs/2015A&A...579A..40S} {579, A40}

\bibitem[\protect\citeauthoryear{{Smith}}{{Smith}}{2014}]{2014ARA&A..52..487S}
{Smith} N.,  2014, \mn@doi [\araa] {10.1146/annurev-astro-081913-040025}, \href {https://ui.adsabs.harvard.edu/abs/2014ARA&A..52..487S} {52, 487}

\bibitem[\protect\citeauthoryear{{Soderberg}, {Chevalier}, {Kulkarni}  \& {Frail}}{{Soderberg} et~al.}{2006}]{2006ApJ...651.1005S}
{Soderberg} A.~M.,  {Chevalier} R.~A.,  {Kulkarni} S.~R.,   {Frail} D.~A.,  2006, \mn@doi [\apj] {10.1086/507571}, \href {https://ui.adsabs.harvard.edu/abs/2006ApJ...651.1005S} {651, 1005}

\bibitem[\protect\citeauthoryear{{Soderberg} et~al.,}{{Soderberg} et~al.}{2010}]{2010Natur.463..513S}
{Soderberg} A.~M.,  et~al., 2010, \mn@doi [\nat] {10.1038/nature08714}, \href {https://ui.adsabs.harvard.edu/abs/2010Natur.463..513S} {463, 513}

\bibitem[\protect\citeauthoryear{{Stockdale} et~al.,}{{Stockdale} et~al.}{2009}]{2009CBET.1714....1S}
{Stockdale} C.~J.,  et~al., 2009, Central Bureau Electronic Telegrams, \href {https://ui.adsabs.harvard.edu/abs/2009CBET.1714....1S} {1714}

\bibitem[\protect\citeauthoryear{{Stroh} et~al.,}{{Stroh} et~al.}{2021}]{2021ApJ...923L..24S}
{Stroh} M.~C.,  et~al., 2021, \mn@doi [\apjl] {10.3847/2041-8213/ac375e}, \href {https://ui.adsabs.harvard.edu/abs/2021ApJ...923L..24S} {923, L24}

\bibitem[\protect\citeauthoryear{{Terreran} et~al.,}{{Terreran} et~al.}{2019}]{2019ApJ...883..147T}
{Terreran} G.,  et~al., 2019, \mn@doi [\apj] {10.3847/1538-4357/ab3e37}, \href {https://ui.adsabs.harvard.edu/abs/2019ApJ...883..147T} {883, 147}

\bibitem[\protect\citeauthoryear{{Valenti}, {Tartaglia}, {Sand}, {Wyatt}, {Bostroem}, {Reichart}, {Haislip}  \& {Kouprianov}}{{Valenti} et~al.}{2017}]{2017ATel10706....1V}
{Valenti} S.,  {Tartaglia} L.,  {Sand} D.,  {Wyatt} S.,  {Bostroem} K.~A.,  {Reichart} D.~E.,  {Haislip} J.~B.,   {Kouprianov} V.,  2017, The Astronomer's Telegram, \href {https://ui.adsabs.harvard.edu/abs/2017ATel10706....1V} {10706}

\bibitem[\protect\citeauthoryear{{Weiler}, {Panagia}, {Montes}  \& {Sramek}}{{Weiler} et~al.}{2002}]{2002ARA&A..40..387W}
{Weiler} K.~W.,  {Panagia} N.,  {Montes} M.~J.,   {Sramek} R.~A.,  2002, \mn@doi [\araa] {10.1146/annurev.astro.40.060401.093744}, \href {https://ui.adsabs.harvard.edu/abs/2002ARA&A..40..387W} {40, 387}

\bibitem[\protect\citeauthoryear{{Wellons}, {Soderberg}  \& {Chevalier}}{{Wellons} et~al.}{2012}]{2012ApJ...752...17W}
{Wellons} S.,  {Soderberg} A.~M.,   {Chevalier} R.~A.,  2012, \mn@doi [\apj] {10.1088/0004-637X/752/1/17}, \href {https://ui.adsabs.harvard.edu/abs/2012ApJ...752...17W} {752, 17}

\bibitem[\protect\citeauthoryear{{Wenger} et~al.,}{{Wenger} et~al.}{2000}]{2000A&AS..143....9W}
{Wenger} M.,  et~al., 2000, \mn@doi [\aaps] {10.1051/aas:2000332}, \href {https://ui.adsabs.harvard.edu/abs/2000A&AS..143....9W} {143, 9}

\bibitem[\protect\citeauthoryear{{Whiting}}{{Whiting}}{2012}]{2012MNRAS.421.3242W}
{Whiting} M.~T.,  2012, \mn@doi [\mnras] {10.1111/j.1365-2966.2012.20548.x}, \href {https://ui.adsabs.harvard.edu/abs/2012MNRAS.421.3242W} {421, 3242}

\bibitem[\protect\citeauthoryear{{Wilson} et~al.,}{{Wilson} et~al.}{2011}]{2011MNRAS.416..832W}
{Wilson} W.~E.,  et~al., 2011, \mn@doi [\mnras] {10.1111/j.1365-2966.2011.19054.x}, \href {https://ui.adsabs.harvard.edu/abs/2011MNRAS.416..832W} {416, 832}

\bibitem[\protect\citeauthoryear{{Wood-Vasey}, {Aldering}, {Nugent}  \& {Chassagne}}{{Wood-Vasey} et~al.}{2003}]{2003IAUC.8082....1W}
{Wood-Vasey} W.~M.,  {Aldering} G.,  {Nugent} P.,   {Chassagne} R.,  2003, \iaucirc, \href {https://ui.adsabs.harvard.edu/abs/2003IAUC.8082....1W} {8082}

\bibitem[\protect\citeauthoryear{{Yaron} \& {Gal-Yam}}{{Yaron} \& {Gal-Yam}}{2012}]{2012PASP..124..668Y}
{Yaron} O.,  {Gal-Yam} A.,  2012, \mn@doi [\pasp] {10.1086/666656}, \href {https://ui.adsabs.harvard.edu/abs/2012PASP..124..668Y} {124, 668}

\bibitem[\protect\citeauthoryear{{Yin} \& {Heeschen}}{{Yin} \& {Heeschen}}{1991}]{1991Natur.354..130Y}
{Yin} Q.~F.,  {Heeschen} D.~S.,  1991, \mn@doi [\nat] {10.1038/354130a0}, \href {https://ui.adsabs.harvard.edu/abs/1991Natur.354..130Y} {354, 130}

\bibitem[\protect\citeauthoryear{Zwart et~al.,}{Zwart et~al.}{2008}]{zwart_2008}
Zwart J. T.~L.,  et~al., 2008, \mn@doi [\mnras] {10.1111/j.1365-2966.2008.13953.x}, 391, 1545

\makeatother
\end{thebibliography}

\clearpage
\appendix
\pagebreak


\section{\textbf{ASKAP Sensitivity Limit}}
\label{subsec: Appendix A}
To obtain an upper limit on the redshift for detectable CCSNe with ASKAP we calculated the sensitivity distance horizon using Mrk 297A, which has the highest known peak radio luminosity for a SN \citep[][and references therein]{1991Natur.354..130Y}. For the VAST-P1 RMS limit $\sigma_{\rm RMS}=0.25$\,mJy beam$^{-1}$, the $5\sigma$ \texttt{Selavy} threshold results in a distance limit of:

\begin{align}
    D=\sqrt\frac{3\times L_{\nu}}{4\pi\times 5\sigma_{\rm RMS}}\approx\frac{3\times 2.43\times 10^{29}}{4\pi \times 1.25} \approx 697.65 \text{ Mpc}  
\end{align}
corresponding to a  redshift of $z\approx 0.146$.
For this calculation we have taken three times the radio luminosity $L_{\nu}\approx2.43\times10^{29}$\,erg s$^{-1}$Hz$^{-1}$, which was obtained by scaling the $12.2$\,mJy flux density at $5$\,GHz to $887.5$\,MHz with an assumed spectral index of $\alpha=-0.7$. This figure may be used only as a conservative limit on the highest possible SN radio luminosity  \citep{2015Perez-Torres}.

\section{\textbf{Full Candidate List }}
\label{subsec: Appendix B}
In Table \ref{tab:Top29Candidates} we provide a complete list of the $29$ CCSNe  late-time radio re-brightening candidates that we identified in the ASKAP data as part of this work. The type, discovery date, and distance are taken from the OpenSNe catalogue.

\begin{table}
\caption{CCSNe identified with ASKAP detections in RACS-low and VAST-P1, arranged chronologically by discovery date.}
\setlength{\tabcolsep}{2pt}
\begin{tabular}{llllcc}
\hline
 Name            & Type      & RA           & Dec           &   Dist. [Mpc] & Disc. [MJD]    \\
\hline
 SN 1978K        & SN II     & 03:17:38.6   & $-$66:33:04   &   4.1         & 43720          \\
 SN 1987A        & SN IIp    & 05:35:27.9   & $-$69:16:11   &   0.054       & 46850          \\
 SN 1996aq       & SN Ic     & 14:22:22.7   & $-$00:23:23   &   23.9        & 50312          \\
 SN 2002cj       & SN Ic     & 15:21:20.7   & $-$19:51:24   &   68          & 52385          \\
 SN 2002hy       & SN Ic     & 10:54:39.2   & $-$21:03:41   &   45          & 52590          \\
 SN 2003bg       & SN IIb    & 04:10:59.4   & $-$31:24:49   &   19.5        & 52695          \\
 SN 2004dk       & SN Ib     & 16:21:48.9   & $-$02:16:17   &   22.6        & 53216          \\
 SN 2004gg       & SN II     & 09:46:48.7   & $+$16:02:46   &   87          & 53325          \\
 SN 2004Q        & SN II     & 12:47:39.7   & $-$26:12:22   &   108         & 53034          \\
 SN 2006bl       & SN II     & 15:39:50.9   & $+$14:11:16   &   145         & 53829          \\
 SN 2006O        & SN II     & 01:01:20.9   & $+$31:30:11   &   66          & 53757          \\
 SN 2007rw       & SN II     & 12:38:03.6   & $-$02:15:41   &   38.3        & 54433          \\
 SN 2008bu       & SN II     & 16:47:24.3   & $-$20:08:33   &   57          & 54573          \\
 SN 2008de       & SN II     & 15:55:24.9   & $-$09:41:47   &   180         & 54573          \\
 SN 2008fi       & SN IIb    & 01:53:23.3   & $+$29:21:31   &   110.8       & 54704          \\
 SN 2012ap       & SN Ic-BL  & 05:00:13.7   & $-$03:20:50   &   53.3        & 55967          \\
 SN 2012dy       & SN II     & 21:18:50.7   & $-$57:38:42   &   45.3        & 56142          \\
 SN 2013bi       & SN IIp    & 18:25:02.1   & $+$27:31:53   &   71.7        & 56375          \\
 SN 2014C        & SN Ib     & 22:37:05.7   & $+$34:24:30   &   12          & 56662          \\
 SN 2015co       & SN II     & 17:14:36.0   & $+$30:44:07   &   131         & 57144          \\
 SN 2016coi      & SN Ic-BL  & 21:59:04.1   & $+$18:11:11   &   16.18       & 57535          \\
 SN 2017ggi      & SN Ib     & 00:58:52.9   & $-$35:51:46   &   178         & 57977          \\
 SN 2017gmr      & SN II     & 02:35:30.2   & $-$09:21:14   &   21.7        & 58000          \\
 SN 2017hyh      & SN IIb    & 07:10:41.0   & $+$06:27:41   &   53.7        & 58062          \\
 SN 2018dhp      & SN II     & 21:32:51.7   & $-$25:20:20   &   141.4       & 58309          \\
 SN 2018keq      & SN Ic-BL  & 23:22:41.8   & $+$21:00:42   &   180         & 58469          \\
 SN 2019arl      & SN Ic     & 14:47:02.0   & $+$11:39:16   &   133.7       & 58522          \\
 SN 2019atg      & SN Ic     & 12:02:57.9   & $-$30:07:51   &   201         & 58521          \\
 SN 2019cac      & SN IIn    & 13:50:43.9   & $-$02:30:23   &   213.8       & 58526          \\

\hline
\end{tabular}

\label{tab:Top29Candidates}
\end{table}

\section{\textbf{Constants from the Literature}}
\label{subsec: Appendix D}

The constant $c_1$ is given (in cgs units) explicitly by \cite{1998ApJ...499..810C}, who cite $c_5$ and $c_6$ from \cite{1970ranp.book.....P}:

\begin{align}
c_1=6.27\times10^{18} \rm
\\
c_5= \frac{\sqrt{3}}{16\pi}\frac{e^3}{mc^2}\left(\frac{\gamma+\frac{7}{3}}{\gamma+1}\right)\Gamma\left(\frac{3\gamma-1}{12}\right)\Gamma\left(\frac{3\gamma+7}{12}\right)
\\
c_6= \frac{\sqrt{3}\pi}{72}em^5c^{10}\left(\gamma+\frac{10}{3}\right)\Gamma\left(\frac{3\gamma+2}{12}\right)\Gamma\left(\frac{3\gamma+10}{12}\right)
\end{align}\\

\section{\textbf{MCMC Corner Plots}}
\label{subsec: Appendix F}

\begin{figure*}
\centering
\includegraphics[width=\textwidth]{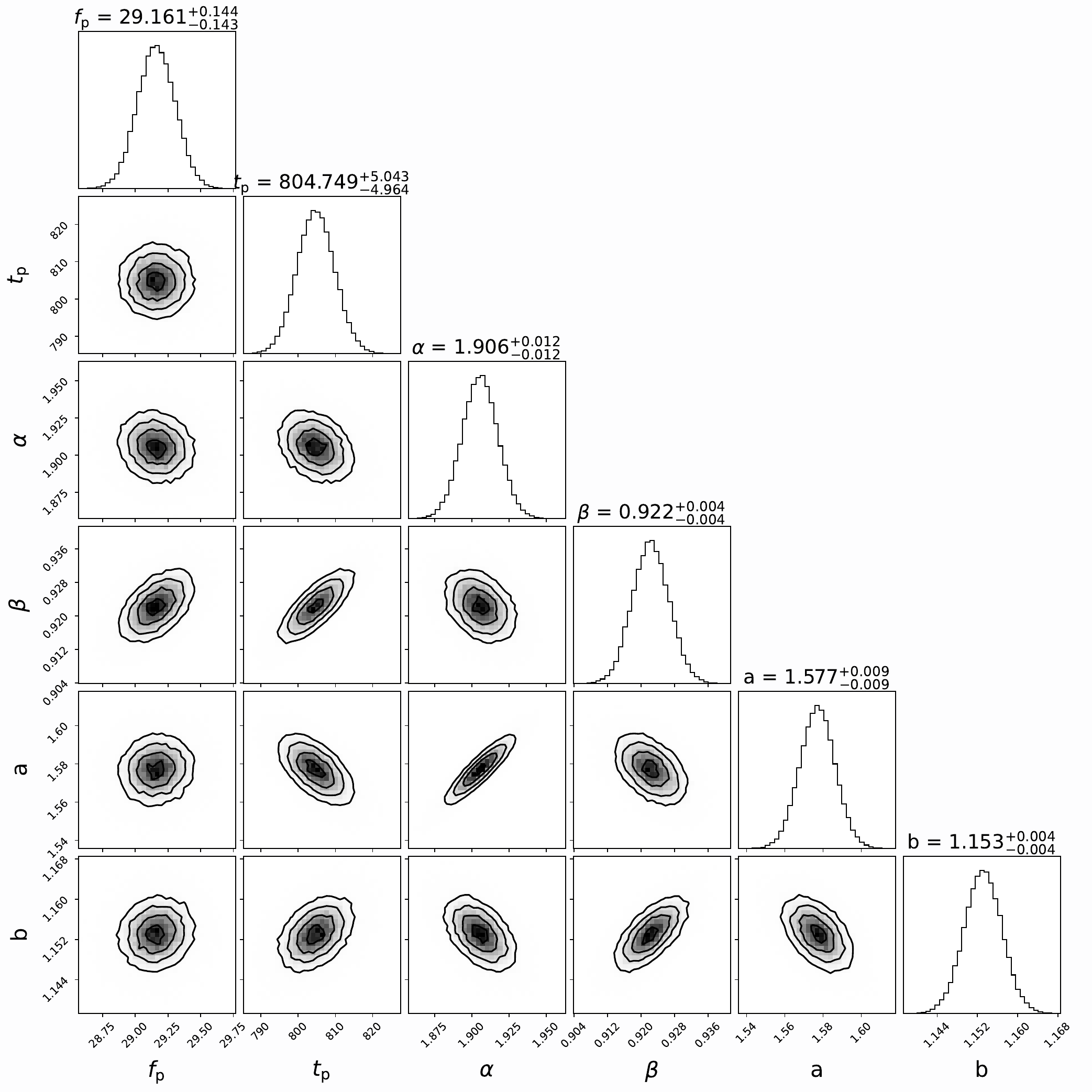}
\caption{SN\,2003bg corner plot of the posterior distributions for the values given in Table \ref{tab:FitParameters}. These are obtained from our \texttt{emcee} sampling and fitting of the early-time ($\Delta t<10^3$\,days) radio data to the parameterised model given in Equation \ref{eq: parameterised model}. The grey colour scale corresponds to the density of samples.}
\label{fig: SN 2003bg Corner Plot}
\end{figure*}

\begin{figure*}
\centering
\includegraphics[width=\textwidth]{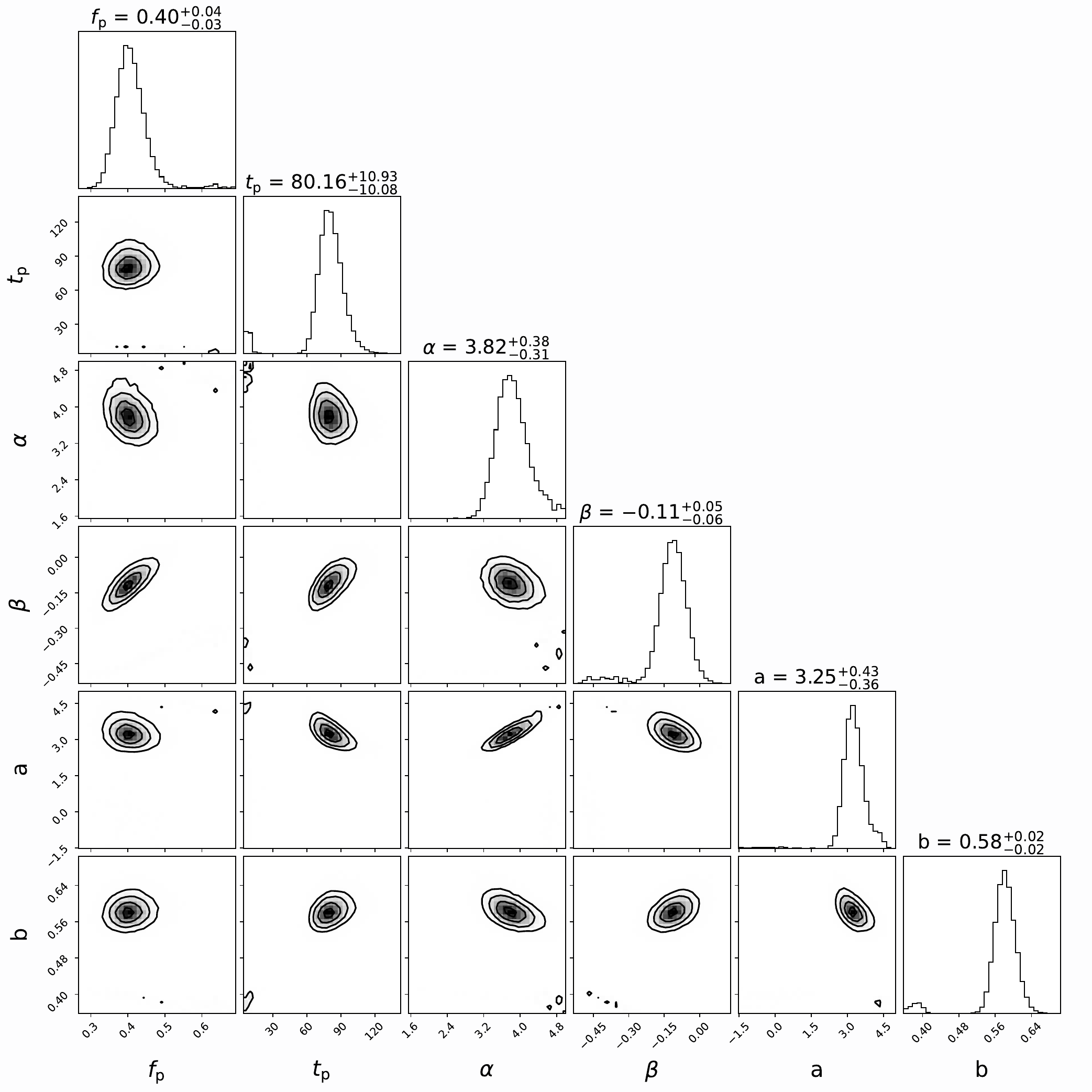}
\caption{SN\,2004dk corner plot of the posterior distributions for the values given in Table \ref{tab:FitParameters}. These are obtained from our \texttt{emcee} sampling and fitting of the early-time ($\Delta t<3\times10^3$\,days) radio data to the parameterised model given in Equation \ref{eq: parameterised model}. The grey colour scale corresponds to the density of samples.} 
\label{fig: SN 2004dk Corner Plot}
\end{figure*}

\begin{figure*}
\centering
\includegraphics[width=\textwidth]{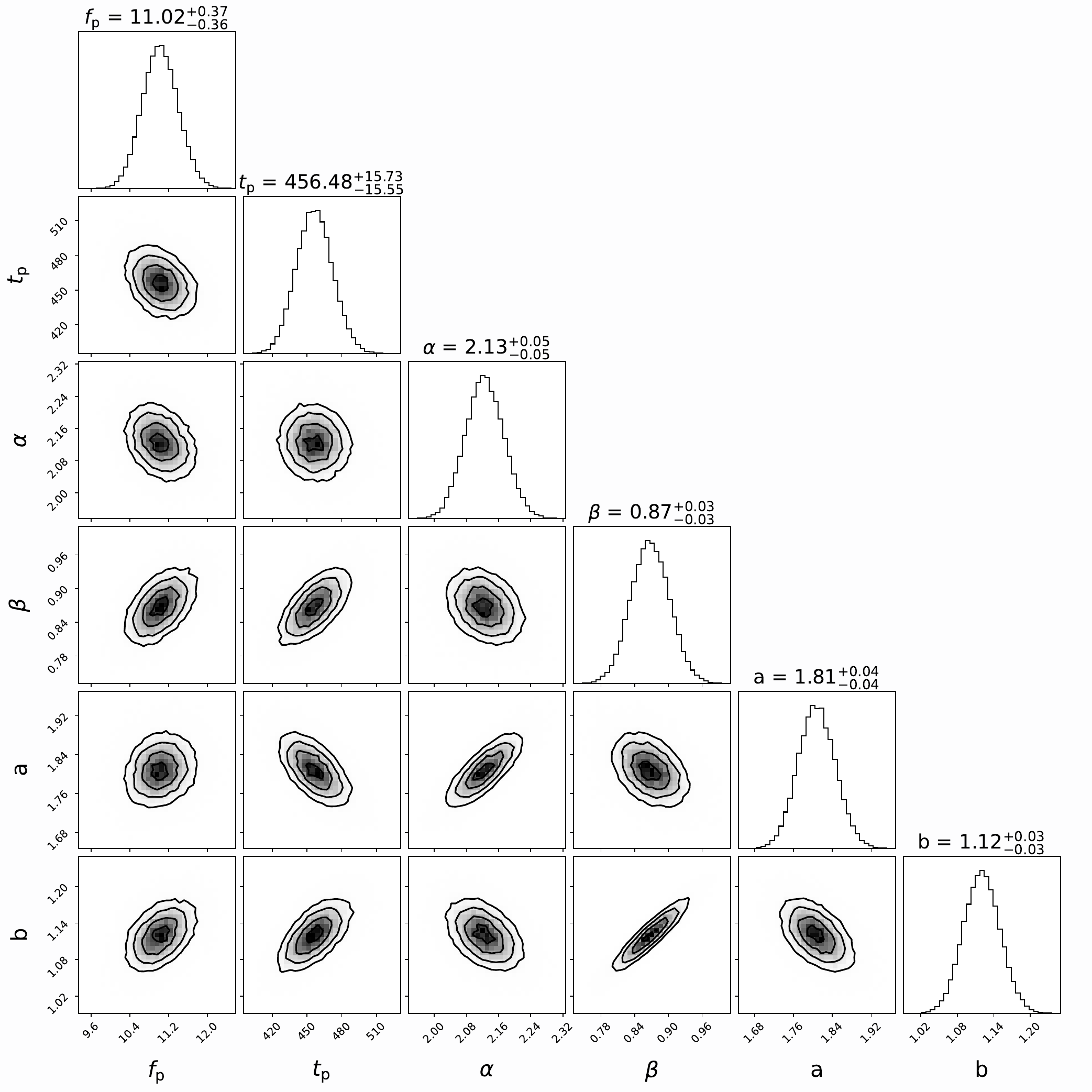}
\caption{SN\,2016coi corner plot of the posterior distributions for the values given in Table \ref{tab:FitParameters}. These are obtained from our \texttt{emcee} sampling and fitting of the early-time ($\Delta t<3\times10^3$\,days) radio data to the parameterised model given in Equation \ref{eq: parameterised model}. The grey colour scale corresponds to the density of samples.} 
\label{fig: SN 2016coi Corner Plot}
\end{figure*}

We provide the corner plots of the posterior distributions for our \texttt{emcee} fitting of SN\,2003bg (Figure \ref{fig: SN 2003bg Corner Plot}), SN\,2004dk (Figure \ref{fig: SN 2004dk Corner Plot}), and SN\,2016coi (Figure \ref{fig: SN 2016coi Corner Plot}). We fit the multi-frequency early-time data to Equation \ref{eq: parameterised model} and assume a spectral peak frequency at the VAST observing frequency $887.5$\,MHz. These plots used the default option for $0.5, 1, 1.5$, and $ 2$ sigma level contours, which respectively contain $11.8$ per cent, $39.3$  per cent, $67.5$ per cent, and $86.4$  per cent of the samples.



\bsp	
\label{lastpage}
\end{document}